\documentclass[journal=jctcce,manuscript=article]{achemso}

\usepackage[version=3]{mhchem} 
\usepackage{graphicx}
\usepackage{amsmath}
\usepackage{verbatim}
\usepackage{url}
\usepackage{geometry}
\usepackage{lscape}
\usepackage{subfigure}

\author{Keyao Pan}
\affiliation[Departments of Bioengineering]
{Departments of Bioengineering and Physics \& Astronomy, Rice University, Houston, Texas 77005}
\author{Michael W. Deem}
\email{mwdeem@rice.edu}
\fax{713-348-5811}
\affiliation[Departments of Bioengineering]
{Departments of Bioengineering and Physics \& Astronomy, Rice University, Houston, Texas 77005}
\altaffiliation{Departments of Physics \& Astronomy}

\title[Free energy calculation for H3N2 influenza]
{Predicting Fixation Tendencies of the H3N2 Influenza Virus by Free Energy Calculation}

\begin{document}
\begin{abstract}
Influenza virus evolves to escape from immune system antibodies that bind to it. We used free energy calculations with Einstein crystals as reference states to calculate the difference of antibody binding free energy ($\Delta\Delta G$) induced by amino acid substitution at each position in epitope B of the H3N2 influenza hemagglutinin, the key target for antibody. A substitution with positive $\Delta\Delta G$ value decreases the antibody binding constant. On average an uncharged to charged amino acid substitution generates the highest $\Delta\Delta G$ values. Also on average, substitutions between small amino acids generate $\Delta\Delta G$ values near to zero. The 21 sites in epitope B have varying expected free energy differences for a random substitution. Historical amino acid substitutions in epitope B for the A/Aichi/2/1968 strain of influenza A show that most fixed and temporarily circulating substitutions generate positive $\Delta\Delta G$ values. We propose that the observed pattern of H3N2 virus evolution is affected by the free energy landscape, the mapping from the free energy landscape to virus fitness landscape, and random genetic drift of the virus. Monte Carlo simulations of virus evolution are presented to support this view.
\end{abstract}

\section{Introduction}
\label{sec:Introduction}

Influenza A virus causes annual global epidemics resulting in 5--15\% of the population being infected, 3--5 million severe cases, and 250,000--500,000 fatalities \cite{WHO2009}. The subtype of influenza A is determined by two surface glycoproteins---hemagglutinin (H) and neuraminidase (N). The H3N2 virus has been one of the dominant circulating subtypes since its emergence in 1968. The antibodies IgG and IgA are the major components of the immune system that control influenza infection, binding to the influenza hemagglutinin \cite{Janeway2005}. There are five epitopes at the antibody binding sites on the top of H3 hemagglutinin, namely epitopes A--E. The epitope bound most prolifically by antibody is defined as the dominant epitope, and it is central to the process of virus neutralization by antibody and virus escape substitution \cite{Gupta2006}. The cellular immune system, on the other hand, plays a relatively less recognized role in handling the invasive influenza virus \cite{Janeway2005}. The cellular system along with the innate immune system exerts a somewhat more homogeneous immune reaction against genetically distinct influenza strains \cite{Lee2008,Janeway2005}.


Vaccination is currently the primary method to prevent and control an influenza epidemic in the human population \cite{WHO2009}. Influenza vaccination raises the level of antibody specific for hemagglutinin and significantly enhances the binding affinity between antibody and hemagglutinin.  Vaccine effectiveness depends on the antigenic distance between the hemagglutinin of the administered vaccine strain and that of the dominant circulating strain in the same season \cite{Gupta2006,Pan2011b}. Memory immune response from virus in previous seasons as well as vaccination in the current and previous seasons impose selective pressure on the current circulating virus to force it to evolve away from the virus strains recognized by memory antibodies that selectively bind to hemagglutinin.

As a result of the immune pressure and the escape evolution of the influenza virus, which is largely substitution in the dominant epitope of hemagglutinin, the influenza vaccine must be redesigned and administered each year, and the vaccine effectiveness has been suboptimal in some flu seasons \cite{Gupta2006,Pan2009}. The escape evolution in the dominant epitope is at a higher rate than that in the amino acid sites outside the dominant epitope \cite{Ferguson2003}. Sites in the dominant epitope also show higher Shannon entropy of the 20 amino acids than do those outside the dominant epitope \cite{Deem2009}. High substitution rate and Shannon entropy in the dominant epitope of hemagglutinin suggest that the dominant epitope is under the strongest positive selection by human antibodies. The immune pressure against each genotype of the dominant epitope can be at least partially quantified by the binding constant between antibody and hemagglutinin.

The H3N2 virus and human immune system in this work are simplified to be a system consisting of the H3 hemagglutinin and the corresponding human antibody. Exposure by infection or vaccination produces an affinity-matured antibody with the binding constant to the corresponding hemagglutinin equal to $10^6$--$10^7$ M$^{-1}$, while the binding constant of an antibody uncorrelated to the hemagglutinin is below $10^2$ M$^{-1}$ \cite{Janeway2005}. Escape substitutions may decrease the binding constant by changing the antibody binding free energy $\Delta G$. Some substitutions decrease the antibody binding constant more than others and have higher probabilities to be fixed, because decrease in the antibody binding constant is favorable to the virus. Here we define the difference of antibody binding free energy as $\Delta\Delta G = \Delta G_{42} - \Delta G_{31}$ in which $\Delta G_{31}$ and $\Delta G_{42}$ are antibody-wildtype hemagglutinin binding free energy and antibody-evolved hemagglutinin binding free energy, respectively, as shown in \ref{fig:scheme}. The fixation tendency of each substitution is a function of the difference of the antibody binding free energy \cite{Zhou2008} of the escape substitution.

Epitope A or B of the H3N2 virus was dominant in most influenza seasons \cite{Gupta2006}. Epitope B of the H3N2 virus was the dominant epitope presenting more substitutions than any other epitope in the recent years. Epitope B was also dominant in 1968 when H3N2 virus emerged. Thus during these periods of time, the substitutions in epitope B directly affect the antibody binding constant and reflect the direction of the virus escape substitution. To attain a global view of the effects of substitutions in epitope B, it is necessary to compute a matrix containing the differences of antibody binding free energy caused by each possible single substitution in epitope B. There are 21 amino acid sites in epitope B, and each residue in the wild type strain may substitute to any of the 19 different types to amino acid residues, hence we need to calculate a $19 \times 21$ matrix with 399 elements. Such a matrix is a free energy landscape quantifying the immune selection over each evolved influenza strain. In this free energy landscape, the virus tends to evolve to a position with low binding affinity of antibody to evade antibodies and reduce the immune pressure. Calculation of this landscape will enable us to study the mechanism of immune escape from a quantitative viewpoint, providing a criterion to describe and foresee the evolution of influenza virus.

This paper is organized as follows: In Materials and Methods section, we describe the protocol for the free energy calculation and the system of hemagglutinin and antibody. In Results section, we present and analyze the calculated free energy landscape. The substitutions observed in history are also compared with the results of the calculation. In the Discussion section, a general picture of H3N2 virus evolution under the selection pressure of the immune system is discussed and simulation results are discussed. Finally, our work is summarized in the Conclusion section.

\section{Materials and Methods}
\label{sec:Materials_and_Methods}

\subsection{Scheme of the Free Energy Calculation}
\label{sec:Scheme_of_the_Free_Energy_Calculation}

\begin{figure}
\centering
\includegraphics[width=6in]{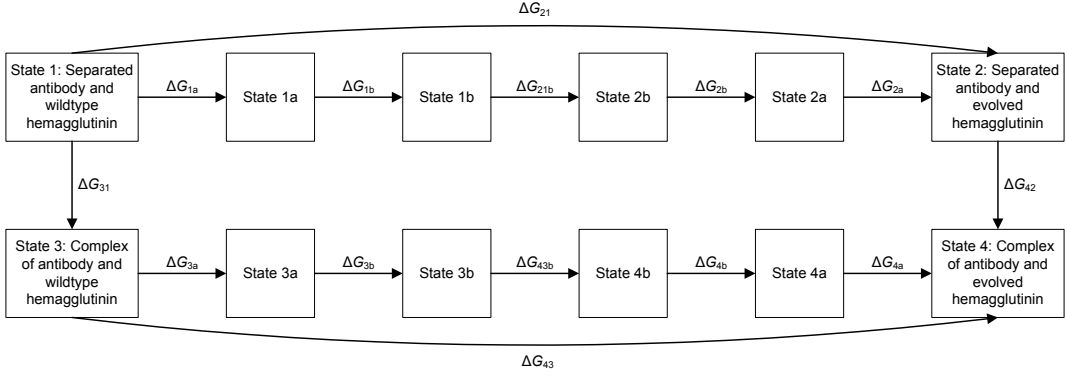}
\caption{The scheme of the free energy calculation. The free energy difference of one substitution is calculated by $\Delta\Delta G = \Delta G_{43} - \Delta G_{21}$. State $n$, $n=1$--$4$, is the real system. State $n$a has the same configuration of atoms as state $n$ except that all the hydrogen atoms have mass 16.000 amu. Compared to state $n$a, state $n$b contains one additional Einstein crystal of product atoms ($n=1,3$) or reactant atoms ($n=2,4$). The mass of hydrogen atoms in state $n$b is also 16.000 amu. Free energy $\Delta G_\mathrm{21b}$ and $\Delta G_\mathrm{43b}$ are obtained by thermodynamic integration.}
\label{fig:scheme}
\end{figure}

The expression of the binding constant $K$ depends on the antibody binding free energy $\Delta G$, $K = \exp\left(-\Delta G/RT\right)$. The Boltzmann constant $R=1.987 \times 10^{-3}$ kcal/mol/K. The temperature is fixed to the normal human body temperature $T = 310$ K. Shown in \ref{fig:scheme}, one substitution in hemagglutinin changes the antibody binding free energy from $\Delta G_{31}$ to $\Delta G_{42}$. The first and second subscripts define the end state and the starting state of the binding process, respectively. The ratio of the antibody binding constant after and before substitution is written as
\begin{equation}\label{eq:ratioK}
\frac{K_1}{K_0} = \exp\left(-\Delta\Delta G / RT\right)
\end{equation}
where $K_1$ and $K_0$ are the antibody binding constant to substituted and wildtype hemagglutinin, respectively.

The difference of the antibody binding free energy $\Delta\Delta G = \Delta G_{42} - \Delta G_{31} = \Delta G_{43} - \Delta G_{21}$ is calculated by applying the Hess' Law to the thermodynamic cycle defined by State 1--4 in \ref{fig:scheme}. The processes corresponding to $\Delta G_{43}$ and $\Delta G_{21}$ are unphysical but more convenient to simulate. We calculated $\Delta G_{21}$ and $\Delta G_{43}$ for each amino acid substitution in the unbound hemagglutinin and hemagglutinin bound by antibody, respectively. On the surface of the virus particle, hemagglutinin exists in the form of a trimer in which three monomers are encoded by the same virus gene. Thus we simultaneously substituted the amino acids in three hemagglutinin monomers in the trimer. The antibody has a Y-shaped structure with two heavy chains and two light chains. In the resolved structure (PDB code: 1KEN), the hemagglutinin trimer is bound by two Fab fragments. Thus, we incorporated the Fab dimer into the system for MD simulation.

Using the software CHARMM \cite{Brooks1983}, we calculated each of $\Delta G_{21}$ and $\Delta G_{43}$ using thermodynamic integration \cite{Frenkel2002}. We used molecular dynamics (MD) simulation to obtain the ensemble averages of the integrand from which each of $\Delta G_{21}$ and $\Delta G_{43}$ is calculated. The potential energy for the MD algorithm to sample the conformation space of the system is
\begin{equation}\label{eq:potential}
U\left(\boldsymbol{r}, \lambda\right) = \left(1-\lambda\right) U_\mathrm{reac}\left(\boldsymbol{r}\right) + \lambda U_\mathrm{prod}\left(\boldsymbol{r}\right)
\end{equation}
in which $\boldsymbol{r}$ is the coordinates of all the atoms, $\lambda$ is the variable of integration, $U_\mathrm{reac}$ is the potential energy of the system corresponding to wildtype hemagglutinin, and $U_\mathrm{prod}$ is the potential energy of the system corresponding to substituted hemagglutinin. The value of $\Delta G_{21}$ or $\Delta G_{43}$ is
\begin{equation}\label{eq:ti}
\Delta G = \int_0^1 \left\langle\frac{\partial U\left(\boldsymbol{r}, \lambda\right)}{\partial\lambda}\right\rangle_\lambda \mathrm{d}\lambda = \int_0^1 \left\langle U_\mathrm{prod}\left(\boldsymbol{r}\right) - U_\mathrm{reac}\left(\boldsymbol{r}\right) \right\rangle_\lambda \mathrm{d}\lambda.
\end{equation}
The integrand $\left\langle U_\mathrm{prod}\left(\boldsymbol{r}\right) - U_\mathrm{reac}\left(\boldsymbol{r}\right) \right\rangle_\lambda$ is the ensemble average with fixed $\lambda$ of potential energy difference between the system after and before substitution. The interval of integration $\lambda \in \left(0,1\right)$ was equally divided into four subintervals in each of which a 16-point Gauss-Legendre quadrature was applied to numerically integrate the ensemble averages. The ensemble averages with 64 distinct $\lambda \in \left(0,1\right)$ were calculated by MD simulation with the potential energy defined in \ref{eq:potential}.

\subsection{Einstein Crystal}
\label{sec:Einstein_Crystal}

We introduce the Einstein crystals to calculate the free energy of the reference state in the dual topology at both endpoints of the thermodynamic integration. To illustrate the function of the Einstein crystals, we analyze the free energy of the dual topology without Einstein crystals when $\lambda = 0$ as an example. We denoted by $n_1$, $n_2$, and $n_0$ numbers of the reactant atoms, product atoms, and all the remaining atoms in the system, respectively. We denoted by $\boldsymbol{r}$, $\boldsymbol{r}_\mathrm{product}$, and $\boldsymbol{x}$ the coordinates of the reactant atoms, product atoms, and all the remaining atoms in the system. The momenta of reactant atoms, product atoms, and the remaining atoms are denoted by $p_{\mathrm{r},i}$, $p_{\mathrm{p},i}$, and $p_{\mathrm{x},i}$. The masses are similarly denoted by $m_{\mathrm{r},i}$, $m_{\mathrm{p},i}$, and $m_{\mathrm{x},i}$. The Hamiltonian of the system with $\lambda = 0$ is
\begin{equation}\label{eq:dual_topology_H}
H = \sum_{i=1}^{n_0}\frac{p_{\mathrm{x},i}^2}{2 m_{\mathrm{x},i}} + \sum_{i=1}^{n_1}\frac{p_{\mathrm{r},i}^2}{2 m_{\mathrm{r},i}} + \sum_{i=1}^{n_2}\frac{p_{\mathrm{p},i}^2}{2 m_{\mathrm{p},i}} + U_{n_0}\left(\boldsymbol{x}\right) + \left(1 - \lambda\right) U_{n_0+n_1}\left(\boldsymbol{x},\boldsymbol{r}\right) + \lambda U_{n_0+n_2}\left(\boldsymbol{x},\boldsymbol{r}_\mathrm{product}\right).
\end{equation}
The partition function is
\begin{align}\label{eq:dual_topology_Q}
Q =& \prod_{i=1}^{n_0} \left(\frac{2\pi m_{\mathrm{x},i}}{h^2 \beta}\right)^{3/2} \prod_{i=1}^{n_1} \left(\frac{2\pi m_{\mathrm{r},i}}{h^2 \beta}\right)^{3/2} \prod_{i=1}^{n_2} \left(\frac{2\pi m_{\mathrm{p},i}}{h^2 \beta}\right)^{3/2} \nonumber\\
& \int \mathrm{d}\boldsymbol{x} \ \mathrm{d}\boldsymbol{r} \ \exp\left[-\beta U_{n_0}\left(\boldsymbol{x}\right) -\beta U_{n_0+n_1}\left(\boldsymbol{x},\boldsymbol{r}\right)\right] \times \int \mathrm{d}\boldsymbol{r}_\mathrm{product} \ 1\nonumber\\
=& Q_\mathrm{real} \times \prod_{i=1}^{n_2} \left[\left(\frac{2\pi m_{\mathrm{p},i}}{h^2 \beta}\right)^{3/2} V\right] \nonumber\\
=& Q_\mathrm{real} \times Q_\mathrm{product}.
\end{align}
When $\lambda = 0$, this partition function is the product of $Q_\mathrm{real}$, the partition function of the real system without product atoms, and $Q_\mathrm{product}$, the partition function of the product atoms when $\lambda = 0$.

The free energy is given by $-1/\beta$ times the logarithm of the above partition function. The free energy is
\begin{equation}\label{eq:dual_topology_G}
G = G_\mathrm{real} - \frac{1}{\beta} \sum_{i=1}^{n_2} \frac{3}{2} \log\left(\frac{2\pi m_{\mathrm{p},i}}{h^2 \beta}\right) - \frac{1}{\beta} n_2 \log V.
\end{equation}
As shown in the above equation, the effect on the translational entropy from the product atoms is proportional to the logarithm of system size $V$. It diverges in the thermodynamic limit. This divergence exists, no matter what $\lambda$ scaling is performed. Note that we do not use the Einstein crystals to handle the translational entropy a ligand loses or gains when binding a flexible biomolecular receptor, which is taken into account by the thermodynamic cycle in \ref{fig:scheme}. The translational entropy, proportional to $\log V$ in \ref{eq:dual_topology_G}, is that of the dummy product atoms, not that of the bound or unbound complex.

The value of $G$ depends on the identity of the product atoms.  Thus, the contribution to the thermodynamic integration is different at the two endpoints, i.e.\ $-kT \log Q_\mathrm{reactant} \neq -kT \log Q_\mathrm{product}$, in which $Q_\mathrm{reactant}$ is the partition function of the reactant atoms when $\lambda = 1$. Note also that the expression of the partition function contains the factor $Q_\mathrm{product}$ for the product atoms. Relating the conventional expression for thermodynamic integration, \ref{eq:ti}, to $\Delta\Delta G$ of \ref{eq:ratioK} requires one to account for this term. This term arises from the use of a dual topology in CHARMM, and this term is typically ignored. While the contribution from the decoupled atoms is not constant, it can be exactly calculated if the restricted partition function over the decoupled atoms can be calculated.  This calculation is what the Einstein crystal performs, using an Einstein crystal for the reference state rather the ideal gas in \ref{eq:dual_topology_H}.

In four 16-window thermodynamic integrations, the smallest variable of integration is $\lambda = 1.32 \times 10^{-3}$. Since $\lambda$ is close to zero, product atoms in the system have potential energy near zero and behave as ideal gas atoms, with translational entropy proportional to the logarithm of system size, see \ref{eq:dual_topology_G}. Exact calculation of the translational entropy terms of product atoms at $\lambda=0$ by explicit dynamics seems difficult, because the translational entropy of the product atoms grows as the logarithm of the system size. These relatively free product atoms destabilize the system. This entropy divergence is a fundamental feature of the statistical mechanics, not a numerical artifact. Unrestrained product atoms induce large fluctuation of the Hamiltonian in the MD algorithm. These fluctuations increase the standard error of the quantity $U_\mathrm{prod}\left(r\right) - U_\mathrm{reac}\left(r\right)$, which is defined in \ref{eq:ti} and is computed from the trajectory of the MD simulation. These fluctuations often cause the numerical integration algorithm in the MD simulation to be unstable \cite{Beveridge1989}. In this case, the energy of the simulated system increases rapidly. This phenomenon causes CHARMM to terminate abnormally. The translational entropy introduced by the free atoms at $\lambda=0$ and 1 affects the result. Reactant atoms cause the same problem near $\lambda=1$.

We noticed that the non-linear scaling, i.e. using a high power of $\lambda$ such as the fourth power of $\lambda$, in \ref{eq:potential} \cite{Mezei1986,Cross1986} did not work. The high power of the smallest $\lambda$ is extremely close to zero and the product atoms are almost free, which cause the MD simulation to terminate abnormally at several windows with small $\lambda$. Additionally, the issue of translational entropy of reactant and product atoms needs to be addressed. Even when the MD algorithm with the non-linear scaling of $\lambda$ \cite{Mezei1986,Cross1986} terminates and appears to have generated a converged simulation trajectory, this does not necessarily imply that the translational entropy of reactant or product atoms has been properly controlled. In fact, the $\lambda$ scaling approach may hide the entropy divergence at $\lambda=0$ or $\lambda=1$ by letting the algorithm terminate due to numerical roundoff error, rather than building statistical mechanical reference states for each of $\lambda=0$ and $\lambda=1$ to account for or control the effect of translational entropy.

An alternative to $\lambda$ scaling introduces the soft-core potential as a way to turn off the potential \cite{Beutler1994,Zacharias1994}. The soft-core approach, like the lambda-scaling approach, does not address the translation entropy of the atoms at $\lambda$ = 0 or $\lambda$ = 1. Previous studies with non-constrained atoms at both endpoints have been performed \cite{Boresch1999,Boresch1999b,Roux1996,Nina1997,Essex1997,Price2001,Zacharias1993}. Besides the classical molecular dynamics with a non-ideal-gas reference state introduced into the dual topology, quantum molecular dynamics via metadynamics has been used to analyze a deamidation process \cite{Kaliman2010}. Other applications of quantum molecular dynamics based free energy calculation include chorismate conversion to prephenate \cite{Crespo2005}, isomerization of glycine \cite{Takahashi2005}, and histone lysine methylation \cite{Wang2007}. As illustrated in \ref{eq:dual_topology_G}, the translational entropy of the uncoupled atoms causes error in the final free energy results if it is not accounted for.

One way to calculate the free energy change exactly is to use a non-ideal-gas reference state. This is quite natural, since the protein is not composed of ideal gas atoms. Deng and Roux introduced restraint potentials to confine the translational and rotational motion of a bound ligand to accelerate convergence of the simulation \cite{Deng2006}. We use this idea to exactly include the contribution from the restrained states and built two Einstein crystals as the reference states for reactant and product atoms, respectively. Our calculation allows a theoretically exact determination of the free energy due to amino acid substitution. 

To handle these two difficulties at both endpoints of the integration in a theoretically exact way, we use two Einstein crystals as the reference states for reactant and product atoms, respectively. The Einstein crystal has been used as a reference state for free energy calculations. Frenkel and Ladd computed free energy of solids by building a path connecting the real solid and the reference Einstein crystal \cite{Frenkel1984}. Noya et al.\ showed that a restrained Einstein crystal is a suitable reference in the free energy calculation of biomolecules \cite{Noya2008}. The Einstein crystal, a solid state model, is consistent with the nature of antibody binding process in liquid phase. First, although the importance of biomolecular flexibility in protein-protein binding process is well-accepted, and is fully and exactly included in our calculation, we simply need to localize the product atoms when $\lambda = 0$ and the reactant atoms when $\lambda = 1$. Moreover, we need to calculate the contribution to the free energy of these localized atoms.

The choice of Einstein crystals as the reference states removes the singularity in thermodynamic integration in \ref{eq:ti}. As an example, an Einstein crystal was used as the reference state for the free energy calculation of hard-sphere fluid in order to remove the singularity in \ref{eq:ti} at the end point $\lambda = 0$ \cite{Frenkel2002b}. In this example, the reference Einstein crystal was achieved by harmonically coupling the particles to their equilibrium positions and removing all interactions between particles \cite{Meijer1990}.

We here use Einstein crystals as the reference states to calculate the binding free energy change due to amino acid substitution. The Einstein crystal is a model for localized atoms. The free energy of the Einstein crystal can be exactly calculated. One Einstein crystal contains distinguishable and non-interacting atoms under harmonic constraints around reference positions fixed in space. In the Einstein crystal, the atom $i$ with coordinates $\boldsymbol{r}_i$ has potential energy
\begin{equation}\label{eq:atomi}
U_i\left(\boldsymbol{r}_i\right) = \frac{K_i}{2}\left\|\boldsymbol{r}_i - \boldsymbol{r}_{i0}\right\|^2
\end{equation}
in which $\boldsymbol{r}_i$ and $\boldsymbol{r}_{i0}$ are the actual and reference position of the atom, respectively, and $K_i$ is the force constant of the harmonic constraint. We denote by $m_i$ the mass of atom $i$. The canonical partition function of an Einstein crystal is
\begin{eqnarray}\label{eq:partition_Einstein}
Q_\mathrm{E}\left(N,V,T\right) &=& \frac{1}{h^{3N}} \int \exp\left(\sum_{i=1}^N \frac{-\beta \boldsymbol{p}_i^2}{2m_i}\right) \exp\left(\sum_{i=1}^N \frac{-\beta K_i \left\|\boldsymbol{r}_i - \boldsymbol{r}_{i0}\right\|^2}{2}\right) \mathrm{d}\boldsymbol{p} \ \mathrm{d}\boldsymbol{r}\nonumber\\
&=& \left(\frac{2\pi}{h\beta}\right)^{3N} \prod_{i=1}^N \left(\frac{m_i}{K_i}\right)^{3/2}.
\end{eqnarray}
The spatial fluctuation of atom $i$ in the Einstein crystal is
\begin{equation}\label{eq:fluctuation}
\left\langle\left(\delta\boldsymbol{r}_{i}\right)^2\right\rangle = \frac{3}{\beta K_i}.
\end{equation}

In our system, we let the potential energy for MD simulation defined by \ref{eq:potential} become
\begin{equation}\label{eq:potential_Einstein}
U\left(\boldsymbol{r},\lambda\right) = \left(1-\lambda\right) U_\mathrm{reac}\left(\boldsymbol{r}\right) + \lambda U_\mathrm{prod}\left(\boldsymbol{r}\right) + \lambda U_\mathrm{ein,reac}\left(\boldsymbol{r}\right) + \left(1-\lambda\right) U_\mathrm{ein,prod}\left(\boldsymbol{r}\right).
\end{equation}
Therefore reactant and product atoms are localized at both $\lambda=0$ and $\lambda=1$. The reference positions of atoms in Einstein crystals are the equilibrium positions of corresponding reactant and product atoms. To minimize the numerical error during the thermodynamic integration calculation, we minimized the fluctuation of the integrand of thermodynamic integration $\left\langle \partial U\left(\boldsymbol{r},\lambda\right) / \partial\lambda\right\rangle_\lambda = \left\langle U_\mathrm{ein,reac}\left(\boldsymbol{r}\right) - U_\mathrm{reac}\left(\boldsymbol{r}\right)\right\rangle_\lambda + \left\langle U_\mathrm{prod}\left(\boldsymbol{r}\right) - U_\mathrm{ein,prod}\left(\boldsymbol{r}\right)\right\rangle_\lambda$. Minimization of the terms on the right hand size is approximately achieved by letting the average spatial fluctuation of each atom in Einstein crystals equal to that of the corresponding reactant or product atom, i.e.
\begin{equation}
\label{eq:fluc_reac}
\left\langle\left(\delta\boldsymbol{r}_{i}\right)^2\right\rangle_\mathrm{reac} = \left\langle\left(\delta\boldsymbol{r}_{i}\right)^2\right\rangle_\mathrm{ein,reac} = \frac{3}{\beta K^\mathrm{reac}_i}
\end{equation}
\begin{equation}
\label{eq:fluc_prod}
\left\langle\left(\delta\boldsymbol{r}_{i}\right)^2\right\rangle_\mathrm{prod} = \left\langle\left(\delta\boldsymbol{r}_{i}\right)^2\right\rangle_\mathrm{ein,prod} = \frac{3}{\beta K^\mathrm{prod}_i}
\end{equation}
For each atom in the Einstein crystal, the force constant of harmonic constraint, $K^\mathrm{reac}_i$ or $K^\mathrm{prod}_i$, was calculated from the monitored fluctuations of the corresponding reactant or product atom with \ref{eq:fluc_reac} or \ref{eq:fluc_prod}. In the scheme in \ref{fig:scheme}, the states with Einstein crystals are states 1b, 2b, 3b, and 4b.

\subsection{Modified Hydrogen Atoms}
\label{sec:Modified_Hydrogen_Atoms}

The frequency of atom vibration depends on its mass. Hydrogen atoms generally have the highest vibration frequencies in the system. Such high frequencies require short time step in MD simulation and increase computational load. To limit vibration frequencies and allow a longer time step, one can apply the SHAKE algorithm to fix the length of any bond involving hydrogen atoms \cite{Ryckaert1977}. The SHAKE algorithm decreases the degrees of freedom in the system by introducing additional constraints between atoms. Instead, we artificially changed the mass of hydrogen atoms from 1.008 to 16.000 amu in order to preserve degree of freedom in the system following the suggestion by Bennett \cite{Bennett1975}. A larger mass of hydrogen atoms allows a longer time step in the MD algorithm. Pomes and McCammon showed that changing the hydrogen mass to 10 amu allow using a 0.01 ps time step to simulate a system which consists of 215 TIP3P water molecules, smaller than our system \cite{Pomes1990}. Feenstra et al.\ change the mass of hydrogen atoms to 4 amu to increase the simulation stability of a system which contains protein and water molecules and resembles our system \cite{Feenstra1999}. We set the time step as 0.001 ps, a value widely used in simulations with physical masses for all atoms, to gain higher stability in the simulation of our large system with a hemagglutinin trimer, a Fab dimer, and water molecules. As with the Einstein crystals, we exactly calculated and subtracted off the contribution of the change to the hydrogen mass to $\Delta\Delta G$. Note that the modification of hydrogen mass is independent to the reference states in the simulation, which is selected to be Einstein crystals in this project. In fact, most of the hydrogen atoms in the system are neither reactant nor product atoms. In \ref{fig:scheme}, the states with Einstein crystals and modified hydrogen atoms are states 1a, 2a, 3a, 4a, 1b, 2b, 3b, and 4b.

\subsection{Expressions of Free Energies}
\label{sec:Expressions_of_Free_Energies}

Introducing two Einstein crystals and heavier hydrogen atoms changes the potential energy in the system, as well as the canonical partition functions. After modification of hydrogen atoms, the mass of atoms changed from $m_{\mathrm{r},i}$ to $m'_{\mathrm{r},i}$, from $m_{\mathrm{p},i}$ to $m'_{\mathrm{p},i}$, or from $m_{\mathrm{x},i}$ to $m'_{\mathrm{x},i}$. Canonical partition functions of the states in \ref{fig:scheme} are:
\begin{align}
\label{eq:state_1}
Q_3\left(n_0+n_1, V, T\right) =& \frac{1}{h^{3\left(n_0+n_1\right)}} \prod_{i=1}^{n_0} \left(\frac{2\pi m_{\mathrm{x},i}}{\beta}\right)^{3/2} \prod_{i=1}^{n_1} \left(\frac{2\pi m_{\mathrm{r},i}}{\beta}\right)^{3/2}\nonumber\\
& \times Z_3\left(n_0+n_1, V, T\right)\\
\label{eq:state_1a}
Q_\mathrm{3a} \left(n_0+n_1, V, T\right) =& \frac{1}{h^{3\left(n_0+n_1\right)}} \prod_{i=1}^{n_0} \left(\frac{2\pi m'_{\mathrm{x},i}}{\beta}\right)^{3/2} \prod_{i=1}^{n_1} \left(\frac{2\pi m'_{\mathrm{r},i}}{\beta}\right)^{3/2}\nonumber\\
& \times Z_3\left(n_0+n_1, V, T\right)\\
\label{eq:state_1b}
Q_\mathrm{3b} \left(n_0+n_1+n_2, V, T\right) =& \frac{1}{h^{3\left(n_0+n_1\right)}} \prod_{i=1}^{n_0} \left(\frac{2\pi m'_{\mathrm{x},i}}{\beta}\right)^{3/2} \prod_{i=1}^{n_1} \left(\frac{2\pi m'_{\mathrm{r},i}}{\beta}\right)^{3/2}\nonumber\\
& \times Z_3\left(n_0+n_1, V, T\right)\left(\frac{2\pi}{h\beta}\right)^{3n_2} \prod_{i=1}^{n_2}\left(\frac{m'_{\mathrm{p},i}}{K^\mathrm{prod}_i}\right)^{3/2}\\
\label{eq:state_2}
Q_4\left(n_0+n_2, V, T\right) =& \frac{1}{h^{3\left(n_0+n_2\right)}} \prod_{i=1}^{n_0} \left(\frac{2\pi m_{\mathrm{x},i}}{\beta}\right)^{3/2} \prod_{i=1}^{n_2} \left(\frac{2\pi m_{\mathrm{p},i}}{\beta}\right)^{3/2}\nonumber\\
& \times Z_4\left(n_0+n_2, V, T\right)\\
\label{eq:state_2a}
Q_\mathrm{4a} \left(n_0+n_2, V, T\right) =& \frac{1}{h^{3\left(n_0+n_2\right)}} \prod_{i=1}^{n_0} \left(\frac{2\pi m'_{\mathrm{x},i}}{\beta}\right)^{3/2} \prod_{i=1}^{n_2} \left(\frac{2\pi m'_{\mathrm{p},i}}{\beta}\right)^{3/2}\nonumber\\
& \times Z_4\left(n_0+n_2, V, T\right)\\
\label{eq:state_2b}
Q_\mathrm{4b} \left(n_0+n_1+n_2, V, T\right) =& \frac{1}{h^{3\left(n_0+n_2\right)}} \prod_{i=1}^{n_0} \left(\frac{2\pi m'_{\mathrm{x},i}}{\beta}\right)^{3/2} \prod_{i=1}^{n_2} \left(\frac{2\pi m'_{\mathrm{p},i}}{\beta}\right)^{3/2}\nonumber\\
& \times Z_4\left(n_0+n_2, V, T\right) \left(\frac{2\pi}{h\beta}\right)^{3n_1} \prod_{i=1}^{n_1}\left(\frac{m'_{\mathrm{r},i}}{K^\mathrm{reac}_i}\right)^{3/2}
\end{align}
in which the states are denoted by the subscripts. Contribution of the potential energy part of the Hamiltonian to the partition function is
\begin{eqnarray}
\label{eq:Z_1}
Z_3\left(n_0+n_1, V, T\right) &=& \int \exp\left(-\beta U_{n_0+n_1}\left(\boldsymbol{r}\right)\right) \mathrm{d}\boldsymbol{r}\\
\label{eq:Z_2}
Z_4\left(n_0+n_2, V, T\right) &=& \int \exp\left(-\beta U_{n_0+n_2}\left(\boldsymbol{r}\right)\right) \mathrm{d}\boldsymbol{r}
\end{eqnarray}
From the partition functions, free energies defined in \ref{fig:scheme} are calculated:
\begin{eqnarray}
\label{eq:dG_1a}
\Delta G_\mathrm{3a} &=& -\frac{3}{2\beta} \sum_{i=1}^{n_0} \ln\left(\frac{m'_{\mathrm{x},i}}{m_{\mathrm{x},i}}\right) -\frac{3}{2\beta} \sum_{i=1}^{n_1} \ln\left(\frac{m'_{\mathrm{r},i}}{m_{\mathrm{r},i}}\right) \\
\label{eq:dG_2a}
\Delta G_\mathrm{4a} &=& \frac{3}{2\beta} \sum_{i=1}^{n_0} \ln\left(\frac{m'_{\mathrm{x},i}}{m_{\mathrm{x},i}}\right) + \frac{3}{2\beta} \sum_{i=1}^{n_2} \ln\left(\frac{m'_{\mathrm{p},i}}{m_{\mathrm{p},i}}\right)\\
\label{eq:dG_1b}
\Delta G_\mathrm{3b} &=& -\frac{3n_2}{\beta} \ln\left(\frac{2\pi}{h\beta}\right) - \frac{3}{2\beta}\sum_{i=1}^{n_2} \ln\left(\frac{m'_{\mathrm{p},i}}{K^\mathrm{prod}_i}\right)\\
\label{eq:dG_2b}
\Delta G_\mathrm{4b} &=& \frac{3n_1}{\beta} \ln\left(\frac{2\pi}{h\beta}\right) + \frac{3}{2\beta}\sum_{i=1}^{n_1} \ln\left(\frac{m'_{\mathrm{r},i}}{K^\mathrm{reac}_i}\right)\\
\label{eq:dG_21b}
\Delta G_\mathrm{43b} &=& -\frac{1}{\beta} \ln\left[\frac{\prod_{i=1}^{n_1}\left(m'_{\mathrm{r},i} / K^\mathrm{reac}_i\right)^{3/2} Z_4\left(n_0+n_2, V, T\right)}{\prod_{i=1}^{n_2}\left(m'_{\mathrm{p},i} / K^\mathrm{prod}_i\right)^{3/2} Z_3\left(n_0+n_1, V, T\right)}\right].
\end{eqnarray}
The free energy between state 3 and 4 is
\begin{equation}\label{eq:dG_21}
\Delta G_{43} = \Delta G_\mathrm{43b} - \frac{1}{\beta} \ln\frac{\left(2\pi / h\beta\right)^{3n_2} \sum_{i=1}^{n_2} \left(m_{\mathrm{p},i} / K^\mathrm{prod}_i\right)^{3/2}}{\left(2\pi / h\beta\right)^{3n_1} \sum_{i=1}^{n_1} \left(m_{\mathrm{r},i} / K^\mathrm{reac}_i\right)^{3/2}} = \Delta G_\mathrm{43b} - \frac{1}{\beta} \ln\frac{Q_\mathrm{E2}\left(n_2,V,T\right)}{Q_\mathrm{E1}\left(n_1,V,T\right)}
\end{equation}
in which $Q_\mathrm{E1}$ and $Q_\mathrm{E2}$ are the partition functions of the Einstein crystals for product atoms and reactant atoms, respectively. The free energy $\Delta G_\mathrm{43b}$ was calculated by thermodynamic integration while $\Delta G_{43}$ was used to calculate the free energy difference of one substitution. Note that the correction term between $\Delta G_\mathrm{43b}$ and $\Delta G_{43}$ is independent of the masses of atoms. Canonical partition functions as well as free energies of the state 1, 1a, 1b, 2, 2a, and 2b are calculated in a similar way.

\subsection{Implementation of Free Energy Calculation Algorithm}
\label{sec:Implementation_of_Free_Energy_Calculation_Algorithm}

The above discussion is the theoretical basis for the implementation of our free energy calculation algorithm. The free energy calculation protocol consists of four steps. First, we built the dual topology with reactant and product atoms in the amino acid substitution site in separated antibody and hemagglutinin or antibody-hemagglutinin complex. We then solvated the protein system and modified the mass of hydrogen atoms. Second, two Einstein crystals were introduced as the reference states for the reactant and product atoms, respectively. Third, the MD simulation was run at 64 windows. The thermodynamic integration algorithm obtained the free energy values $\Delta G_{21}$ for separated antibody and hemagglutinin or $\Delta G_{43}$ for antibody-hemagglutinin complex, as in \ref{fig:scheme}. This step gave the $\Delta\Delta G$ value. Fourth, we calculated the error bar of the $\Delta\Delta G$ value obtained in the last step. The technical details of these four steps are illustrated in the text below. Also described are the verification of the free energy calculation protocol, the software and hardware information, and the CPU hours consumed by the protocol.

The hemagglutinin trimer of H3N2 virus strain A/Aichi/2/1968 with bound dimer antibody HC63 (PDB code: 1KEN) was used in our calculation. For each amino acid substitution, we built the dual topology with side chains of both amino acids prior to the simulation. Reactant and product atoms were defined as the side chains in the original and substituting amino acid, respectively. All the covalent and non-bonded interactions between reactant and product atoms were removed. The protein was in an explicit water box with periodic boundary condition. The mass of hydrogen atoms was changed from 1.008 to 16.000 amu.

All the simulations were performed by CHARMM c33b2 with CHARMM22 force field \cite{Brooks1983}. We first fixed the positions of hemagglutinin trimer except reactant atoms and minimized the system with 200 steps of steepest descent (SD) algorithm and 5000 steps of adopted basis Newton-Raphson (ABNR) algorithm. We ran a 5 ps MD simulation of the system, the trajectory of which gave the spatial fluctuation $\left\langle\left(\delta\boldsymbol{r}_i\right)^2\right\rangle$ of each reactant atom. Then we fixed reactant atoms, released product atoms, and ran a 5 ps MD simulation to obtain the spatial fluctuation of each product atom. Final positions of both reactant and product were adopted as the reference positions of the corresponding Einstein crystal. The force constant $K_i$ of each atom in Einstein crystals was obtained from $\left\langle\left(\delta\boldsymbol{r}_i\right)^2\right\rangle$ by \ref{eq:fluc_reac} and \ref{eq:fluc_prod}. With modified hydrogen atoms and two Einstein crystals as the reference states of reactant and product atoms, state 1b, 2b, 3b, and 4b in \ref{fig:scheme} were generated for thermodynamic integration.

In thermodynamic integration, MD simulations were run at 64 windows with distinct $\lambda$. In each window, pressure of the system was first calibrated with a 10 ps MD simulation in an isothermal-isobaric (NPT) ensemble. The duration of 10 ps is appropriate because it is long enough to equilibrate the pressure and short enough to prevent the protein from drifting away from the original location. We fixed coordinates of the residues and water molecules except for those within 15 \r{A} from the three alpha carbons. Then we removed amino acid residues and water molecules other than those within 27.5 \r{A} from the three alpha carbons of substituted residues in the hemagglutinin trimer to reduce the system size, because the fixed atoms are not included in the topology of movable atoms and the cutoff of the non-bonded forces is 12 \r{A}. The Ewald sum was used to calculate charge interactions. Note that this substantial reduction of the system relies on the assumption that the free energy change due to the amino acid substitution is mostly affected by atoms near the binding site after the system reaches equilibrium. This assumption is based on two facts: the conformations of hemagglutinin and antibody are stable once the system reaches equilibrium, and all the removed or fixed atoms have invariant interactions with the substituting amino acid residues. The stable protein conformation means amino acid residues far away from the substituting residue do not move during the amino acid substitution process. In the CHARMM22 force field used in this project, the cutoff of non-bonded force is 12 \r{A} and less than the 15 \r{A} threshold for system reduction. The system reduction does not directly affect the force on the substituted residue because of absence of the long-range non-bonded force between the substituted residue and atoms removed from the system. This system reduction method was also applied to compute binding free energy of subtilisin \cite{Rao1987}, of tripsin \cite{Essex1997}, and of Src SH2 domain \cite{Price2001}. Robust results were obtained in all of these applications. Generally, this system reduction strategy can produce reliable result if the reduced system contains the residues and molecules critical to the binding process \cite{Essex1997}. We note that the system reduction method could be a limitation of the free energy calculation model. The fixing of amino acid residues and water molecules described in section 2.5 substantially reduced the CPU time needed, but is an approximation to the real system containing the whole proteins. This limitation reflects the tradeoff between model accuracy and required computational resource. In the canonical ensemble, the new system was equilibrated for 200 ps and simulated for another 900 ps as the data production phase. The integrand of thermodynamic integration is the ensemble average of the sampled trajectory $\left\langle \partial U\left(\boldsymbol{r},\lambda\right) / \partial\lambda\right\rangle_\lambda = \left\langle U_\mathrm{ein,reac}\left(\boldsymbol{r}\right) - U_\mathrm{reac}\left(\boldsymbol{r}\right) - U_\mathrm{ein,prod}\left(\boldsymbol{r}\right) + U_\mathrm{prod}\left(\boldsymbol{r}\right)\right\rangle_\lambda$. The free energy $\Delta G_{21}$ and $\Delta G_{43}$ between the real states was calculated by adding a correction term of the Einstein crystals in \ref{eq:dG_21}. Finally, the difference of antibody binding free energy is $\Delta\Delta G = \Delta G_{43} - \Delta G_{21}$.

Error bars of $\Delta\Delta G$ are also given. The convergence behavior of the simulation was analyzed using the block average method developed by Flyvbjerg and Petersen \cite{Flyvbjerg1989}. As mentioned above, the MD simulation for either the unbound hemagglutinin or the hemagglutinin-antibody complex contains 64 windows with distinct $\lambda$. The 900 ps data production phase contains $9 \times 10^5$ simulation steps. The values $A = U_\mathrm{prod}\left(r\right) - U_\mathrm{reac}\left(r\right)$, as in equation \ref{eq:ti}, computed in consecutive simulation steps were grouped into bins, and consecutive bins were merged progressively. The quantity $\sigma^2 \left(A\right) / \left(n-1\right)$, in which $\sigma^2 \left(A\right)$ is the variance of the average of each bin $A_1, A_2, \dots, A_n$ and $n$ is the number of bins, increases with the bin size and reaches a plateau when the bin size is $1 \times 10^4$ steps. We fixed the bin size to $1 \times 10^4$ steps and estimate the variance of ensemble average $\left\langle A \right\rangle$ as $\sigma^2 \left(A\right) / \left(n-1\right)$, following Flyvbjerg and Petersen's method \cite{Flyvbjerg1989}.

This protocol, without the Einstein crystal contribution, was verified by recalculating published free energy differences of amino acid substitution T131I \cite{Zhou2008}. Without the Einstein crystal contribution, our protocol gave the $\Delta\Delta G = 5.69 \pm 0.07$ kcal/mol, compared to the $\Delta\Delta G = 5.20 \pm 0.94$ kcal/mol in the published work \cite{Zhou2008}. Theoretically exact results presented here include the Einstein crystal contribution. We note that the theoretically exact $\Delta\Delta G$ for T131I, including the Einstein crystal contribution, is $3.71 \pm 0.07$ kcal/mol.

The simulation was performed using CHARMM22 force field at three clusters: tg-steele.purdue.teragrid.org (Intel Xeon E5410, 2.33 GHz), sugar.rice.edu (Intel Xeon E5440, 2.83 GHz), and biou.rice.edu (IBM POWER7, 3.55 GHz), as well as at the condor pool tg-condor.rcac.purdue.edu at Purdue University. Simulation of each substitution took approximately 7.5 thousand CPU hours on average, and so this work consumed about three million CPU hours.

\section{Results}
\label{sec:Results}

\subsection{Free Energy Landscape}
\label{sec:Free_Energy_Landscape}

For each of the 21 amino acid sites in epitope B, we substituted from alanine to each one of the 19 other amino acids, in which we used the neutral histidine (CHARMM code: Hse) as the model of histidine. The free energy difference and standard error of each substitution were calculated by the MD simulation (see Materials and Methods). The wildtype amino acid in each site of epitope B was extracted from the hemagglutinin sequence of the H3N2 strain A/Aichi/2/1968. The free energy difference and standard error of the substitution from the wildtype amino acid in each site were then calculated from the values for the change from the wildtype amino acid to alanine and from alanine to the new amino acid. The values are listed in \ref{tab:ddG}.

\begin{table}
\caption{Summary of the calculated free energy differences $\Delta\Delta G$ in each amino acid site in epitope B from the wildtype amino acid to all 20 amino acids. The standard errors are also listed. The free energy difference and its standard error of the substitution from the wildtype amino acid to itself are both zero. The units of free energy differences and their standard errors are kcal/mol.} \centering
{\tiny
\begin{tabular}{l l l l l l l l}
\\ \hline\hline
Positions & 128 & 129 & 155 & 156 & 157 & 158 & 159 \\ \hline
Ala & $-13.12 \pm 0.27$ & $3.33 \pm 0.29$ & $2.78 \pm 0.20$ & $1.19 \pm 0.33$ & $2.48 \pm 0.21$ & $4.27 \pm 0.31$ & $5.18 \pm 0.21$ \\ \hline
Arg & $22.57 \pm 0.46$ & $2.31 \pm 0.45$ & $16.98 \pm 0.37$ & $0.08 \pm 0.50$ & $-4.19 \pm 0.44$ & $-1.61 \pm 0.48$ & $7.07 \pm 0.42$ \\ \hline
Asn & $-4.80 \pm 0.36$ & $5.83 \pm 0.42$ & $-7.83 \pm 0.30$ & $10.72 \pm 0.40$ & $5.64 \pm 0.34$ & $3.41 \pm 0.42$ & $10.97 \pm 0.35$ \\ \hline
Asp & $4.52 \pm 0.38$ & $19.12 \pm 0.42$ & $16.28 \pm 0.32$ & $11.06 \pm 0.42$ & $9.95 \pm 0.37$ & $18.37 \pm 0.40$ & $15.34 \pm 0.36$ \\ \hline
Cys & $-11.83 \pm 0.34$ & $12.64 \pm 0.37$ & $-2.37 \pm 0.30$ & $5.32 \pm 0.38$ & $-2.72 \pm 0.29$ & $-7.88 \pm 0.40$ & $7.92 \pm 0.32$ \\ \hline
Gln & $-12.37 \pm 0.40$ & $7.34 \pm 0.42$ & $-4.29 \pm 0.36$ & $13.14 \pm 0.41$ & $-0.45 \pm 0.36$ & $11.47 \pm 0.43$ & $6.54 \pm 0.40$ \\ \hline
Glu & $11.15 \pm 0.38$ & $10.50 \pm 0.42$ & $17.77 \pm 0.34$ & $26.54 \pm 0.43$ & $4.68 \pm 0.36$ & $8.58 \pm 0.48$ & $5.19 \pm 0.39$ \\ \hline
Gly & $-9.93 \pm 0.39$ & $0.00 \pm 0.00$ & $17.00 \pm 0.34$ & $0.11 \pm 0.44$ & $0.21 \pm 0.36$ & $0.00 \pm 0.00$ & $-4.19 \pm 0.41$ \\ \hline
Hse & $4.43 \pm 0.42$ & $0.15 \pm 0.43$ & $2.47 \pm 0.36$ & $-6.89 \pm 0.43$ & $12.18 \pm 0.38$ & $5.54 \pm 0.46$ & $1.06 \pm 0.39$ \\ \hline
Ile & $-16.03 \pm 0.41$ & $0.54 \pm 0.40$ & $1.55 \pm 0.33$ & $8.33 \pm 0.42$ & $11.22 \pm 0.37$ & $8.09 \pm 0.43$ & $18.96 \pm 0.39$ \\ \hline
Leu & $-23.58 \pm 0.41$ & $-4.27 \pm 0.43$ & $-8.92 \pm 0.33$ & $2.64 \pm 0.45$ & $-6.26 \pm 0.39$ & $1.61 \pm 0.45$ & $4.08 \pm 0.38$ \\ \hline
Lys & $3.57 \pm 0.45$ & $11.18 \pm 0.46$ & $14.58 \pm 0.37$ & $0.00 \pm 0.00$ & $6.24 \pm 0.40$ & $-1.60 \pm 0.48$ & $5.39 \pm 0.46$ \\ \hline
Met & $-13.38 \pm 0.39$ & $-2.59 \pm 0.39$ & $1.23 \pm 0.35$ & $10.11 \pm 0.43$ & $16.15 \pm 0.36$ & $14.49 \pm 0.44$ & $-6.38 \pm 0.37$ \\ \hline
Phe & $-10.21 \pm 0.45$ & $6.12 \pm 0.43$ & $9.39 \pm 0.35$ & $0.30 \pm 0.45$ & $10.28 \pm 0.40$ & $5.17 \pm 0.48$ & $12.33 \pm 0.42$ \\ \hline
Pro & $-9.36 \pm 0.36$ & $-2.43 \pm 0.42$ & $-1.86 \pm 0.31$ & $2.32 \pm 0.43$ & $5.69 \pm 0.30$ & $17.09 \pm 0.40$ & $6.08 \pm 0.36$ \\ \hline
Ser & $-14.55 \pm 0.34$ & $3.36 \pm 0.37$ & $-1.09 \pm 0.29$ & $-1.45 \pm 0.38$ & $0.00 \pm 0.00$ & $2.76 \pm 0.39$ & $0.00 \pm 0.00$ \\ \hline
Thr & $0.00 \pm 0.00$ & $7.35 \pm 0.36$ & $0.00 \pm 0.00$ & $-1.08 \pm 0.41$ & $6.34 \pm 0.32$ & $8.36 \pm 0.41$ & $15.32 \pm 0.32$ \\ \hline
Trp & $9.82 \pm 0.47$ & $4.81 \pm 0.47$ & $19.84 \pm 0.43$ & $23.26 \pm 0.48$ & $16.14 \pm 0.45$ & $3.52 \pm 0.62$ & $-1.35 \pm 0.45$ \\ \hline
Tyr & $-14.83 \pm 0.43$ & $2.72 \pm 0.42$ & $7.25 \pm 0.36$ & $-2.18 \pm 0.46$ & $-8.37 \pm 0.44$ & $18.42 \pm 0.51$ & $5.95 \pm 0.43$ \\ \hline
Val & $-19.13 \pm 0.37$ & $3.56 \pm 0.38$ & $8.57 \pm 0.31$ & $-3.01 \pm 0.39$ & $7.63 \pm 0.32$ & $3.77 \pm 0.42$ & $6.45 \pm 0.32$ \\ \hline
\\ \hline\hline
Positions & 160 & 163 & 165 & 186 & 187 & 188 & 189 \\ \hline
Ala & $4.16 \pm 0.22$ & $-0.24 \pm 0.22$ & $4.15 \pm 0.24$ & $-3.19 \pm 0.19$ & $-4.03 \pm 0.23$ & $3.45 \pm 0.25$ & $-9.01 \pm 0.28$ \\ \hline
Arg & $9.70 \pm 0.44$ & $5.97 \pm 0.39$ & $14.58 \pm 0.41$ & $21.01 \pm 0.38$ & $8.12 \pm 0.42$ & $-0.06 \pm 0.45$ & $-0.39 \pm 0.48$ \\ \hline
Asn & $2.07 \pm 0.34$ & $-2.32 \pm 0.32$ & $0.00 \pm 0.00$ & $4.67 \pm 0.30$ & $-10.07 \pm 0.34$ & $0.00 \pm 0.00$ & $-3.18 \pm 0.37$ \\ \hline
Asp & $13.50 \pm 0.32$ & $12.64 \pm 0.32$ & $25.01 \pm 0.31$ & $24.54 \pm 0.28$ & $7.78 \pm 0.35$ & $19.77 \pm 0.37$ & $6.77 \pm 0.35$ \\ \hline
Cys & $15.82 \pm 0.31$ & $1.84 \pm 0.30$ & $1.93 \pm 0.29$ & $-2.30 \pm 0.25$ & $-11.09 \pm 0.32$ & $4.07 \pm 0.34$ & $6.23 \pm 0.33$ \\ \hline
Gln & $3.04 \pm 0.39$ & $-8.29 \pm 0.35$ & $4.27 \pm 0.36$ & $5.16 \pm 0.33$ & $-2.87 \pm 0.37$ & $12.36 \pm 0.39$ & $0.00 \pm 0.00$ \\ \hline
Glu & $15.48 \pm 0.36$ & $2.17 \pm 0.35$ & $15.74 \pm 0.34$ & $33.29 \pm 0.31$ & $14.41 \pm 0.35$ & $10.10 \pm 0.37$ & $12.16 \pm 0.39$ \\ \hline
Gly & $1.22 \pm 0.39$ & $-5.83 \pm 0.38$ & $9.11 \pm 0.37$ & $0.13 \pm 0.27$ & $-0.60 \pm 0.30$ & $-5.06 \pm 0.32$ & $-5.69 \pm 0.32$ \\ \hline
Hse & $0.52 \pm 0.38$ & $6.31 \pm 0.33$ & $7.44 \pm 0.33$ & $18.15 \pm 0.30$ & $3.69 \pm 0.39$ & $-1.95 \pm 0.40$ & $-8.53 \pm 0.40$ \\ \hline
Ile & $1.51 \pm 0.34$ & $10.62 \pm 0.36$ & $3.85 \pm 0.33$ & $-1.85 \pm 0.30$ & $-2.51 \pm 0.33$ & $-4.77 \pm 0.37$ & $3.65 \pm 0.37$ \\ \hline
Leu & $-1.39 \pm 0.40$ & $3.85 \pm 0.35$ & $-9.20 \pm 0.37$ & $1.07 \pm 0.30$ & $-0.40 \pm 0.38$ & $-1.30 \pm 0.37$ & $-6.91 \pm 0.39$ \\ \hline
Lys & $5.91 \pm 0.44$ & $10.37 \pm 0.38$ & $1.93 \pm 0.41$ & $-1.15 \pm 0.39$ & $24.91 \pm 0.41$ & $8.42 \pm 0.44$ & $9.48 \pm 0.64$ \\ \hline
Met & $10.78 \pm 0.38$ & $7.22 \pm 0.35$ & $1.63 \pm 0.36$ & $13.06 \pm 0.33$ & $-5.11 \pm 0.36$ & $6.97 \pm 0.38$ & $6.86 \pm 0.40$ \\ \hline
Phe & $7.90 \pm 0.41$ & $-0.86 \pm 0.36$ & $13.87 \pm 0.38$ & $6.94 \pm 0.33$ & $-7.23 \pm 0.39$ & $2.05 \pm 0.39$ & $4.37 \pm 0.43$ \\ \hline
Pro & $4.51 \pm 0.32$ & $12.50 \pm 0.34$ & $18.96 \pm 0.33$ & $11.82 \pm 0.29$ & $10.69 \pm 0.32$ & $-10.24 \pm 0.35$ & $-8.98 \pm 0.36$ \\ \hline
Ser & $7.13 \pm 0.29$ & $9.07 \pm 0.30$ & $-0.92 \pm 0.28$ & $0.00 \pm 0.00$ & $-4.88 \pm 0.31$ & $8.09 \pm 0.33$ & $-5.09 \pm 0.34$ \\ \hline
Thr & $0.00 \pm 0.00$ & $9.18 \pm 0.30$ & $10.35 \pm 0.31$ & $-14.79 \pm 0.27$ & $0.00 \pm 0.00$ & $3.53 \pm 0.38$ & $9.30 \pm 0.35$ \\ \hline
Trp & $0.86 \pm 0.44$ & $12.34 \pm 0.35$ & $19.02 \pm 0.43$ & $-7.69 \pm 0.38$ & $-11.04 \pm 0.48$ & $7.20 \pm 0.40$ & $-9.19 \pm 0.45$ \\ \hline
Tyr & $-5.43 \pm 0.39$ & $1.06 \pm 0.34$ & $14.76 \pm 0.37$ & $11.90 \pm 0.33$ & $5.29 \pm 0.42$ & $1.57 \pm 0.40$ & $4.81 \pm 0.41$ \\ \hline
Val & $7.99 \pm 0.34$ & $0.00 \pm 0.00$ & $9.79 \pm 0.32$ & $2.97 \pm 0.29$ & $3.08 \pm 0.33$ & $3.73 \pm 0.34$ & $-7.89 \pm 0.36$ \\ \hline
\\ \hline\hline
Positions & 190 & 192 & 193 & 194 & 196 & 197 & 198 \\ \hline
Ala & $-18.12 \pm 0.24$ & $-0.86 \pm 0.23$ & $-5.20 \pm 0.20$ & $2.37 \pm 0.23$ & $5.95 \pm 0.23$ & $-2.40 \pm 0.29$ & $0.00 \pm 0.00$ \\ \hline
Arg & $4.97 \pm 0.41$ & $23.07 \pm 0.44$ & $32.33 \pm 0.41$ & $-13.66 \pm 0.37$ & $-25.38 \pm 0.44$ & $-17.94 \pm 0.47$ & $3.99 \pm 0.37$ \\ \hline
Asn & $-16.44 \pm 0.30$ & $-2.56 \pm 0.32$ & $8.24 \pm 0.30$ & $-3.81 \pm 0.31$ & $13.27 \pm 0.36$ & $-6.58 \pm 0.38$ & $0.05 \pm 0.28$ \\ \hline
Asp & $18.75 \pm 0.32$ & $2.92 \pm 0.32$ & $15.29 \pm 0.29$ & $26.72 \pm 0.35$ & $9.25 \pm 0.34$ & $5.58 \pm 0.39$ & $5.17 \pm 0.24$ \\ \hline
Cys & $-20.36 \pm 0.32$ & $-1.45 \pm 0.32$ & $-9.79 \pm 0.26$ & $1.91 \pm 0.30$ & $1.30 \pm 0.31$ & $6.70 \pm 0.36$ & $5.91 \pm 0.22$ \\ \hline
Gln & $-17.37 \pm 0.37$ & $-6.00 \pm 0.37$ & $4.87 \pm 0.34$ & $-0.83 \pm 0.32$ & $7.68 \pm 0.36$ & $0.00 \pm 0.00$ & $1.41 \pm 0.31$ \\ \hline
Glu & $0.00 \pm 0.00$ & $1.18 \pm 0.35$ & $45.40 \pm 0.34$ & $38.35 \pm 0.33$ & $3.60 \pm 0.36$ & $11.34 \pm 0.41$ & $2.37 \pm 0.30$ \\ \hline
Gly & $-17.09 \pm 0.29$ & $-13.46 \pm 0.30$ & $-13.89 \pm 0.27$ & $-18.59 \pm 0.30$ & $8.08 \pm 0.31$ & $4.11 \pm 0.36$ & $3.65 \pm 0.28$ \\ \hline
Hse & $-26.26 \pm 0.35$ & $-0.96 \pm 0.38$ & $-0.96 \pm 0.35$ & $9.95 \pm 0.34$ & $18.42 \pm 0.37$ & $-2.62 \pm 0.40$ & $-3.27 \pm 0.35$ \\ \hline
Ile & $-16.45 \pm 0.37$ & $-5.57 \pm 0.37$ & $-3.80 \pm 0.31$ & $-6.91 \pm 0.32$ & $0.77 \pm 0.34$ & $1.23 \pm 0.41$ & $0.01 \pm 0.32$ \\ \hline
Leu & $-17.27 \pm 0.36$ & $-7.97 \pm 0.37$ & $10.76 \pm 0.34$ & $0.00 \pm 0.00$ & $10.07 \pm 0.39$ & $-0.03 \pm 0.40$ & $-11.18 \pm 0.29$ \\ \hline
Lys & $-9.33 \pm 0.38$ & $5.67 \pm 0.42$ & $39.36 \pm 0.39$ & $-16.67 \pm 0.38$ & $0.49 \pm 0.40$ & $-16.50 \pm 0.47$ & $1.98 \pm 0.37$ \\ \hline
Met & $-26.63 \pm 0.34$ & $6.82 \pm 0.36$ & $-2.91 \pm 0.32$ & $7.75 \pm 0.35$ & $4.08 \pm 0.37$ & $-7.79 \pm 0.40$ & $15.57 \pm 0.32$ \\ \hline
Phe & $-31.89 \pm 0.39$ & $1.56 \pm 0.40$ & $16.46 \pm 0.59$ & $2.78 \pm 0.34$ & $-1.99 \pm 0.37$ & $1.05 \pm 0.44$ & $8.73 \pm 0.34$ \\ \hline
Pro & $-17.85 \pm 0.33$ & $-2.28 \pm 0.33$ & $9.84 \pm 0.31$ & $8.01 \pm 0.31$ & $15.42 \pm 0.35$ & $-5.34 \pm 0.40$ & $0.70 \pm 0.29$ \\ \hline
Ser & $-14.75 \pm 0.31$ & $-7.79 \pm 0.30$ & $0.00 \pm 0.00$ & $6.62 \pm 0.29$ & $6.91 \pm 0.29$ & $1.97 \pm 0.36$ & $-2.40 \pm 0.22$ \\ \hline
Thr & $-4.17 \pm 0.32$ & $0.00 \pm 0.00$ & $-2.04 \pm 0.27$ & $12.40 \pm 0.31$ & $7.81 \pm 0.33$ & $-7.91 \pm 0.36$ & $6.79 \pm 0.24$ \\ \hline
Trp & $-22.93 \pm 0.39$ & $2.31 \pm 0.44$ & $17.92 \pm 0.42$ & $-1.30 \pm 0.40$ & $8.17 \pm 0.43$ & $-7.73 \pm 0.44$ & $-7.23 \pm 0.38$ \\ \hline
Tyr & $-13.82 \pm 0.38$ & $7.63 \pm 0.42$ & $16.16 \pm 0.38$ & $9.73 \pm 0.36$ & $2.92 \pm 0.40$ & $6.10 \pm 0.44$ & $-4.82 \pm 0.32$ \\ \hline
Val & $-9.12 \pm 0.31$ & $-6.80 \pm 0.32$ & $-6.92 \pm 0.30$ & $2.59 \pm 0.29$ & $0.00 \pm 0.00$ & $4.16 \pm 0.39$ & $-4.22 \pm 0.24$ \\ \hline
\end{tabular}
} \label{tab:ddG}
\end{table}

As described in \ref{eq:dG_21}, each $\Delta\Delta G$ value listed in \ref{tab:ddG} contains the contribution of two Einstein crystals. The contribution of Einstein crystals to the final $\Delta\Delta G$ values was calculated for each of the 399 amino acid substitutions in epitope B. The average fraction of the contribution of Einstein crystals in the calculated $\Delta\Delta G$ values is 44\%. The contribution of Einstein crystals is far greater than that of the statistical error of our free energy calculation in \ref{tab:ddG}, which is 4.5\% on average. Thus, the Einstein crystal contribution is both theoretically exact and practically important. In 371 of the 399 substitutions, the absolute values of the contribution of Einstein crystals is greater than 1.96 standard errors of the final $\Delta\Delta G$ values. That is, the contribution of Einstein crystals is significant with $p < 0.05$ in 93.0\% of all the amino acid substitutions. Consequently, it is essential to incorporate Einstein crystals in the free energy calculation to eliminate the error caused by the methods that neglect the unknown effect of the translational entropy of the free atoms in thermodynamic integration. The contribution of the translational entropy of ideal gas-like atoms ($\lambda=0$ or $\lambda=1$) needs to be either calculated or removed by a theoretically exact method to perform an exact free energy calculation.

The obtained $\Delta\Delta G$ values allow us to analyze the character of each of the 20 amino acids. We first averaged over all the 21 amino acid sites in epitope B the $\Delta\Delta G$ value caused by the single substitutions from alanine to the other amino acids. The averaged $\Delta\Delta G$ values are listed in \ref{tab:rank}. The largest $\Delta\Delta G$ are caused by the negatively charged amino acids (Glu, Asp) and the positively charged amino acids (Arg, Lys), indicating that introduction of charged amino acids in the dominant epitope decreases the binding affinity between antibody and hemagglutinin. Note that amino acid substitutions that change the charge of hemagglutinin significantly affect the calculated free energy values \cite{Morgan2010,Hunenberger1999,Figueirido1995}. The issue of how to best calculate free energy differences when charge changes has been debated over the years. In the present paper, we are using the standard Ewald approach with explicit solvent. We note that the evolutionary history of H3 hemagglutinin since 1968 shows an increasing trend of the number of charged amino acids in epitope B \cite{Pan2011a}, which agrees with the results that introduction of charged amino facilitates virus evasion from antibody, as illustrated in \ref{tab:rank}. The result that introduction of charged amino acid on average increases $\Delta\Delta G$ is not an artifact, is supported by data from the influenza evolution, and is expected on the basis that charge is hydrophilic. In addition to the charge, the rank of free energy differences also largely correlated to the size of amino acid. By the definition used by RasMol \cite{Sayle1995}, the 16 uncharged amino acids are tagged as hydrophobic (Ala, Gly, Ile, Leu, Met, Phe, Pro, Trp, Tyr, Val), large (Gln, Hse, Ile, Leu, Met, Phe, Trp, Tyr), medium (Asn, Cys, Pro, Thr, Val), and small (Ala, Gly, Ser), as shown in \ref{tab:rank}. The ranks of small amino acids are lower than those of medium amino acids ($p = 0.036$, Wilcoxon rank-sum test) and those of large amino acids ($p = 0.085$, Wilcoxon rank-sum test). In contrast, the hydrophobicity of the uncharged amino acids is largely uncorrelated to their ranks by $\Delta\Delta G$. As a result, charged amino acids in the dominant epitope are essential to the immune evasion while the virus escape substitution among small amino acids have minimal effect.

\begin{table}
\caption{The rank of the average binding free energy difference of the single substitution from alanine to another amino acid over all the 21 amino acid sites in epitope B of hemagglutinin trimer.  The rank correlates with the charge and the size of amino acid, and it is relatively uncorrelated to the hydrophobicity. Here we applied classifications of RasMol for the biochemical properties of the 20 amino acids \cite{Sayle1995}. The relative frequencies of 20 amino acids were counted from the H3 sequences in NCBI database from 1968 to 2009.} \centering
{\scriptsize
\begin{tabular}{l l l c c c c c l}
\\\hline
Rank & Amino Acid & $\Delta\Delta G$ (kcal/mol)    & Charged & Hydrophobic & Large & Medium & Small & Relative frequency\\\hline
1 &		Glu &		$14.612 \pm 0.061$  & $\times$ & & $\times$ & & & 0.029\\
2 &		Asp &		$14.533 \pm 0.055$  & $\times$ & & & $\times$ & & 0.051\\
3 &		Arg &		$6.018 \pm 0.078$   & $\times$ & & $\times$ & & & 0.052\\
4 &		Lys &		$5.766 \pm 0.078$   & $\times$ & & $\times$ & & & 0.057\\
5 &		Trp &		$4.458 \pm 0.081$   & & $\times$ & $\times$ & & & 0.016\\
6 &		Tyr &		$3.984 \pm 0.071$   & & $\times$ & $\times$ & & & 0.035\\
7 &		Thr &		$3.981 \pm 0.050$   & & & & $\times$ & & 0.078\\
8 &		Pro &		$3.912 \pm 0.054$   & & $\times$ & & $\times$ & & 0.060\\
9 &		Met &		$3.562 \pm 0.062$   & & $\times$ & $\times$ & & & 0.009\\
10 &		Phe &		$3.522 \pm 0.073$ & & $\times$ & $\times$ & & & 0.030\\
11 &		Hse &		$2.654 \pm 0.064$ & & & $\times$ & & & 0.020\\
12 &		Gln &		$1.985 \pm 0.063$ & & & $\times$ & & & 0.042\\
13 &		Ile &		$1.396 \pm 0.060$ & & $\times$ & $\times$ & & & 0.070\\
14 &		Asn &		$1.150 \pm 0.054$ & & & & $\times$ & & 0.085\\
15 &		Val &		$1.147 \pm 0.051$ & & $\times$ & & $\times$ & & 0.055\\
16 &		Cys &		$0.888 \pm 0.046$ & & & & $\times$ & & 0.028\\
17 &		Ser &		$0.469 \pm 0.044$ & & & & & $\times$  & 0.096\\
(18) &	(Ala) &		($0.000 \pm 0.000$)   & & $\times$ & & & $\times$  & 0.046\\
19 &		Gly &		$-1.612 \pm 0.055$ & & $\times$ & & & $\times$  & 0.070\\
20 &		Leu &		$-2.273 \pm 0.064$ & & $\times$ & $\times$ & & & 0.071\\
\hline
\end{tabular}
}
\label{tab:rank}
\end{table}

Epitope B comprises 21 amino acid sites in the top of the hemagglutinin trimer. Taking the probability for one substituting amino acid to exist at each site to be proportional to the relative frequency of this amino acid in H3 hemagglutinin, the weighted average free energy difference in each of the 21 sites was calculated. The relative frequencies of 20 amino acids were obtained from 6896 H3 hemagglutinin sequences deposited between 1968 and 2009 in the NCBI database \cite{Pan_select} and listed in \ref{tab:rank}. Also using the $\Delta\Delta G$ values in \ref{tab:ddG}, we calculated and tabulated in \ref{tab:site} for each site $i$ the value of $\left\langle\Delta\Delta G\right\rangle_i$, which is the average $\Delta\Delta G$ weighted by the probability for each different amino acid to be introduced, where probability is proportional to the relative frequencies of 20 amino acids counted from the H3 sequences in NCBI database from 1968 to 2009.

\begin{table}
\caption{The rank of the average free energy difference $\left\langle\Delta\Delta G\right\rangle_i$ generated by a substitution in each amino acid site $i$ of epitope B.} \centering
\begin{tabular}{l l l}
\\\hline
Rank & Site & $\left\langle\Delta\Delta G\right\rangle_i$ (kcal/mol) \\\hline
1 &	193 &	$8.074 \pm 0.081$ \\
2 &	159 &	$7.792 \pm 0.094$ \\
3 &	165 &	$7.741 \pm 0.086$ \\
4 &	158 &	$6.128 \pm 0.108$ \\
5 &	196 &	$5.444 \pm 0.088$ \\
6 &	160 &	$4.956 \pm 0.090$ \\
7 &	186 &	$4.754 \pm 0.076$ \\
8 &	163 &	$4.722 \pm 0.085$ \\
9 &	129 &	$4.690 \pm 0.103$ \\
10 &	155 &	$4.471 \pm 0.081$ \\
11 &	156 &	$4.029 \pm 0.106$ \\
12 &	157 &	$3.944 \pm 0.090$ \\
13 &	188 &	$2.945 \pm 0.092$ \\
14 &	194 &	$1.886 \pm 0.080$ \\
15 &	187 &	$1.182 \pm 0.087$ \\
16 &	198 &	$0.531 \pm 0.072$ \\
17 &	189 &	$-0.631 \pm 0.098$ \\
18 &	192 &	$-1.737 \pm 0.087$ \\
19 &	197 &	$-1.967 \pm 0.099$ \\
20 &	128 &	$-7.746 \pm 0.098$ \\
21 &	190 &	$-12.666 \pm 0.084$ \\
\hline
\end{tabular}
\label{tab:site}
\end{table}

As shown in \ref{tab:site}, there is obvious variation among the expected free energy differences $\left\langle\Delta\Delta G\right\rangle_i$ caused by single substitutions at amino acid site $i$ of epitope B. This variation is partly due to the wildtype amino acids in the sites. For instance, the wildtype amino acid in site 190 is Glu that has the highest rank in \ref{tab:rank}. As shown in \ref{tab:site}, any amino acid substitution in site 190 tends to have a negative $\Delta\Delta G$. Another cause of variation in $\left\langle\Delta\Delta G\right\rangle_i$ is that distinct sites affect differently the antibody binding process. Epitope B of the wildtype A/Aichi/2/1968 hemagglutinin sequence contains five sites with threonine: 128, 155, 160, 187, and 192. The mathematical expectancies $\left\langle\Delta\Delta G\right\rangle_i$ in these five sites are $-7.746$, $4.471$, $4.956$, $1.182$, and $-1.737$ kcal/mol, respectively. Therefore, each site in epitope B has a specific effect on the virus escape substitution. A random substitution in epitope B affects the antibody binding free energy differently depending on the site and the substituting amino acids.

\begin{figure}
\centering
\includegraphics[width=5in]{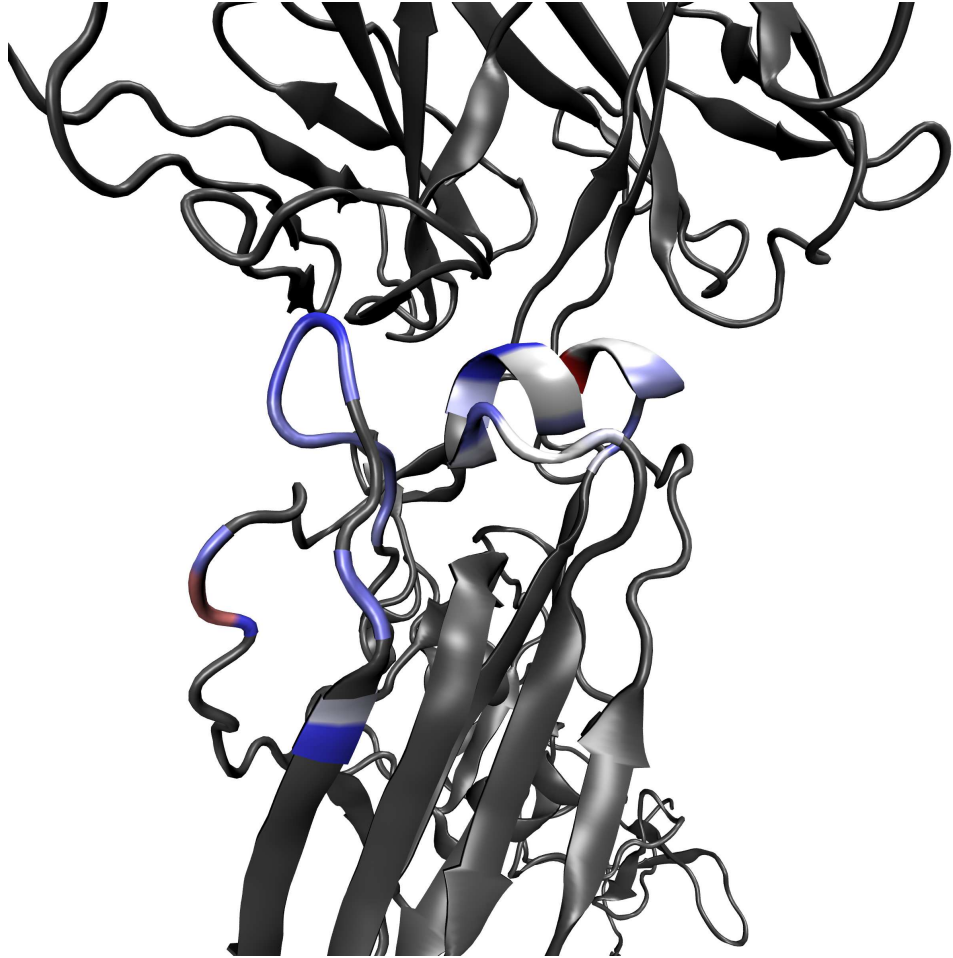}
\caption{The tertiary structure of the interface between the HA1 domain of H3 hemagglutinin monomer A/Aichi/2/1968 (bottom) and the antibody HC63 (top) (PDB code: 1KEN). Water molecules are not shown. Epitope B of the HA1 domain is located in two loops and one $\alpha$-helix with the color scale modulated according to the expected free energy difference $\left\langle\Delta\Delta G\right\rangle_i$ of each site $i$ in epitope B. The color scale ranges from red for the most negative $\left\langle\Delta\Delta G\right\rangle_i$ values to blue for the most positive $\left\langle\Delta\Delta G\right\rangle_i$ values. The sites $i$ in epitope B with $\left\langle\Delta\Delta G\right\rangle_i$ near zero are colored white. The region outside epitope B is colored gray. The red site 128 is far from the antibody binding region and the red site 190 possessed the original amino acid Glu, which is a charged amino acid. It may explain why these two sites show negative $\left\langle\Delta\Delta G\right\rangle_i$ with large absolute values.}
\label{fig:epitope}
\end{figure}

The variation of $\left\langle\Delta\Delta G\right\rangle_i$ is also reflected by the tertiary structure of the epitope B bound by the antibody. By looking into the structure of epitope B shown in \ref{fig:epitope}. Epitope B resides in two protruding loops from amino acid site 128 to 129, and from site 155 to 165, respectively, and in a $\alpha$-helix from site 186 to 198. Site 128 has a negative average free energy difference $\left\langle\Delta\Delta G\right\rangle_{128} = -7.746 \pm 0.098$ kcal/mol. All the other sites in these two loops show a positive $\left\langle\Delta\Delta G\right\rangle_i$ value of a random substitution, with the minimum $\left\langle\Delta\Delta G\right\rangle_{157} = 3.944 \pm 0.090$ kcal/mol in site 157. The $\alpha$-helix is located between hemagglutinin and antibody. In the $\alpha$-helix, the sites facing towards the antibody usually present large positive $\left\langle\Delta\Delta G\right\rangle_i$ values such as site 193 and 196, while the sites facing towards the hemagglutinin show lower $\left\langle\Delta\Delta G\right\rangle_i$ such as site 189, 192, and 197. Thus in the one dimensional sequence from site 186 to 198, the $\left\langle\Delta\Delta G\right\rangle_i$ values oscillate with peaks and valleys corresponding to the sites in the $\alpha$-helix facing alternatingly to the antibody and hemagglutinin. Consequently, the variation of the expected free energy changes in distinct sites depends on the structure of the hemagglutinin-antibody complex.

\subsection{Historical Substitutions in Epitope B}
\label{sec:Historical_Mutations_in_Epitope_B}

The simulation results are supported in two aspects by amino acid sequence data of H3 hemagglutinin collected since 1968. These historical sequences are downloaded from the NCBI Influenza Virus Resource \cite{NCBI_FLU} and aligned. First, Pan et al.\ analyzed the number of charged amino acid in epitope B of H3 hemagglutinin in each year since 1968, and found an increasing trend of charged amino acids \cite{Pan2011a}. This finding supports the results that amino acid substitution introducing charged residues on average facilitates virus escape from antibody, as illustrated in \ref{tab:rank}. Second, amino acid substitutions in epitope B between 1968 and 1975 also verified the free energy calculation, as shown below.

With the knowledge of the free energy landscape of the single substitutions, we are able to recognize favorable single substitutions in epitope B. Substitutions with large positive $\Delta\Delta G$ values enable the virus to evade the immune pressure and increase the virus fitness. Favorable substitutions grow in the virus population. Selection for substitutions with large $\Delta\Delta G$ is part of the evolutionary strategy of the virus. The results of free energy calculation can also explain the substituted virus strains collected in history.

We analyzed the hemagglutinin sequence information of H3N2 strains evolving from the A/Aichi/2/1968 strains. H3 hemagglutinin circulating from 1968 to 1971 was mainly in the HK68 antigenic cluster while those circulating from 1972 to 1975 were mainly in the EN72 antigenic cluster \cite{Smith2004}. \ref{tab:history} shows that in the dominant epitope B, there were 17 substitutions occurred in 12 sites collected between 1968 to 1975 \cite{Shih2007}, which contributed to the immune evasion and corresponding virus evolution from the HK68 cluster to the EN72 cluster. Also listed in \ref{tab:history} are the free energy differences of these historical substitutions. The 17 substituting amino acids have significantly higher ranks compared to the corresponding wildtype amino acids ($p = 0.0044$, Wilcoxon signed-rank test). This significant difference is expected because 15 of 17 substituting amino acids have ranks between 1 and 10, while 10 of 12 wildtype amino acids in the substituted site have ranks between 11 and 20. In all the 21 sites in epitope B, 15 of 21 wildtype amino acids have ranks between 11 and 20. Additionally, the $\Delta\Delta G$ values of these 17 substitutions listed in \ref{tab:history} are greater than the expected free energy differences $\left\langle\Delta\Delta G\right\rangle_i$ in \ref{tab:site} of random substitutions in the 12 substituted sites ($p = 0.013$, Wilcoxon signed-rank test).

\begin{table}
\caption{Substitutions occurred in epitope B of the hemagglutinin A/Aichi/2/1968 (H3N2) as of 1975. Also listed are the time when the substitutions were observed, and the free energy differences with standard errors. In each site of epitope B, all the 20 amino acid were sorted in the descending order by the free energy differences introduced by a substitution from the wildtype amino acid to 20 amino acids. The ranks of the substituting amino acid and the wildtype amino acid in each substituted site are listed in the column Rank (substituting) and Rank (WT), respectively.} \centering
\begin{tabular}{l l l l l}
\\\hline
Substitution & Year & $\Delta\Delta G$ (kcal/mol) & Rank (substituting) & Rank (WT) \\\hline
T128N &	1971       &	$  -4.796 \pm    0.361$ &	 8 &	 7\\
T128I &	1975       &	$ -16.026 \pm    0.412$ &	18 &	 7\\
G129E &	1970, 1972 &	$  10.500 \pm    0.415$ &	 4 &	17\\
T155Y &	1972--1973, fixed in 1973 &	$   7.254 \pm    0.358$ &	 9 &	14\\
G158E &	1971--1972 &	$   8.584 \pm    0.479$ &	 6 &	17\\
S159N &	1971, 1974--1975 &	$  10.969 \pm    0.352$ &	 5 &	17\\
S159C &	1972       &	$   7.923 \pm    0.324$ &	 6 &	17\\
S159R &	1972       &	$   7.065 \pm    0.424$ &	 7 &	17\\
T160A &	1973       &	$   4.160 \pm    0.217$ &	11 &	18\\
S186N &	1975       &	$   4.673 \pm    0.298$ &	10 &	14\\
N188D &	1971--1973, fixed in 1973 &	$  19.767 \pm    0.367$ &	 1 &	14\\
Q189K &	1975       &	$   9.484 \pm    0.640$ &	 2 &	10\\
E190V &	1972       &	$  -9.115 \pm    0.310$ &	 5 &	 3\\
E190D &	1975       &	$  18.752 \pm    0.324$ &	 1 &	 3\\
S193N &	1972--1975 &	$   8.239 \pm    0.301$ &	10 &	12\\
S193D &	1975       &	$  15.285 \pm    0.294$ &	 7 &	12\\
A198T &	1972       &	$   6.793 \pm    0.236$ &	 3 &	14\\
\hline
\end{tabular}
\label{tab:history}
\end{table}

We also looked into the historical escape substitutions in epitope B evading the immune pressure of the vaccine strains. For each influenza season, the amino acids in the administered vaccine strain were defined as the wildtype ones and those in the dominant circulating strain as the substituting amino acids. In each of the 19 seasons in which H3N2 virus was the dominant subtype from 1971 to 2004, the substitutions in epitope B were located \cite{Gupta2006} and their $\Delta\Delta G$ values were obtained from \ref{tab:ddG}. As shown in \ref{tab:vaccine}, escape substitutions in epitope B as of 1973 mostly had positive $\Delta\Delta G$ and generated substituting amino acids with increased rank ($p = 0.047$, Wilcoxon signed-rank test). Such tendency to introduce amino acids with higher ranks was not observed after 1973: the ranks of wildtype and substituting amino acids after 1973 present little significant difference ($p = 0.28$, Wilcoxon signed-rank test). The hemagglutinin of A/Aichi/2/1968 used in the free energy calculating is in the HK68 antigenic cluster. Perhaps after the virus evolved into the next EN72 cluster, change in the virus antigenic character stimulates the immune system to produce new types of antibody other than the HC63 antibody used in the calculation. A different binding antibody changes the free energy landscape of the substitutions in epitope B. Thus the application of the present free energy landscape should be limited within the HK68 and EN72 clusters. Free energy differences of substitutions in the EN72 cluster would need to be calculated using the updated antibody crystal structure.

\begin{table}
\caption{Substitutions occurred in epitope B of H3 hemagglutinin between the vaccine strain and the dominant circulating strain in each season in which the H3N2 subtype was dominant. The free energy difference with standard error of each substitution is obtained using the free energy landscape in \ref{tab:ddG}. The ranks of free energy differences sorted in the descending order are listed in column Rank (vaccine) and in column Rank (circulating) for the amino acids in the vaccine strain and the dominant circulating strain, respectively.} \centering
\begin{tabular}{l l l l l}
\\\hline
Year &	Substitution &	$\Delta\Delta G$ (kcal/mol) &	Rank (vaccine) &	Rank (circulating) \\\hline
1972 &	T155Y &	$   7.254 \pm    0.358$ &	14 &	 9 \\
1972 &	G158E &	$   8.584 \pm    0.479$ &	17 &	 6 \\
1972 &	S159C &	$   7.923 \pm    0.324$ &	17 &	 6 \\
1972 &	E190V &	$  -9.115 \pm    0.310$ &	 3 &	 5 \\
1973 &	T160A &	$   4.160 \pm    0.217$ &	18 &	11 \\
1973 &	N188D &	$  19.767 \pm    0.367$ &	14 &	 1 \\
1973 &	S193N &	$   8.239 \pm    0.301$ &	12 &	10 \\
1975 &	S157L &	$  -6.256 \pm    0.394$ &	15 &	19 \\
1975 &	A160T &	$  -4.160 \pm    0.217$ &	11 &	18 \\
1975 &	Q189K &	$   9.484 \pm    0.640$ &	10 &	 2 \\
1975 &	N193D &	$   7.046 \pm    0.317$ &	10 &	 7 \\
1984 &	E156K &	$ -26.536 \pm    0.429$ &	 1 &	15 \\
1984 &	V163A &	$  -0.243 \pm    0.217$ &	15 &	16 \\
1984 &	D190E &	$ -18.752 \pm    0.324$ &	 1 &	 3 \\
1984 &	I196V &	$  -0.768 \pm    0.343$ &	16 &	18 \\
1987 &	Y155H &	$  -4.782 \pm    0.414$ &	 9 &	11 \\
1987 &	E188D &	$   9.669 \pm    0.382$ &	 3 &	 1 \\
1987 &	K189R &	$  -9.872 \pm    0.697$ &	 2 &	11 \\
1996 &	V190D &	$  27.867 \pm    0.299$ &	 5 &	 1 \\
1996 &	L194I &	$  -6.914 \pm    0.324$ &	13 &	17 \\
1997 &	K156Q &	$  13.140 \pm    0.413$ &	15 &	 3 \\
1997 &	E158K &	$ -10.187 \pm    0.515$ &	 6 &	18 \\
1997 &	V190D &	$  27.867 \pm    0.299$ &	 5 &	 1 \\
1997 &	L194I &	$  -6.914 \pm    0.324$ &	13 &	17 \\
1997 &	V196A &	$   5.947 \pm    0.229$ &	18 &	11 \\
2003 &	H155T &	$  -2.472 \pm    0.355$ &	11 &	14 \\
2003 &	Q156H &	$ -20.028 \pm    0.365$ &	 3 &	20 \\
2003 &	S186G &	$   0.132 \pm    0.275$ &	14 &	13 \\
\hline
\end{tabular}
\label{tab:vaccine}
\end{table}

\section{Discussion}
\label{sec:Discussion}

\subsection{Fitness of the Virus Strains}
\label{sec:Fitness}

The free energy landscape shown in \ref{tab:ddG} gives the change of the antibody binding affinity, $K_1/K_0 = \exp\left(-\Delta\Delta G/RT\right)$, induced by each possible substitution in epitope B of the wildtype hemagglutinin. The majority of the substitutions lead to positive $\Delta\Delta G$, and yield a reduced binding affinity $K_1$ that is smaller than the binding affinity of the original mature antibody $K_0$. Decreased antibody binding constant grants the virus a higher chance of evading the immune pressure and infecting host cells. We propose that virus fitness is positively correlated to the free energy difference $\Delta\Delta G$. The other factor affecting virus fitness is the capability of the hemagglutinin to maintain the normal biochemical functions, such as virus entry. Most sites in epitope B changed amino acid identities during 1968 to 2005 as the H3N2 virus kept circulating \cite{Shih2007}. We therefore postulate that the substitutions in epitope B do not greatly interfere with the biochemical function of hemagglutinin, and virus fitness is dominantly determined by the free energy difference resulted from substitutions in epitope B.

The binding constant between hemagglutinin and antibody after the first round of maturation is about $10^6$ M$^{-1}$, and the binding constant of an uncorrelated antibody is below $10^2$ M$^{-1}$ \cite{Janeway2005}. On average, four substitutions in epitope B change the substituted hemagglutinin sufficiently so that the immune response of the original antibody binding to epitope B is abrogated \cite{Gupta2006}. Since this is a reduction of the binding constant from roughly $10^6$ M$^{-1}$ to $10^2$ M$^{-1}$, one amino acid substitution that contributes to immune escape causes on average a 10-fold decrease in antibody binding constant, or equivalently $\Delta\Delta G_\mathrm{crit} = 1.42$ kcal/mol at 310 K. Assuming the effect of immune evasion can be broken into the sum of individual amino acid substitutions in the dominant epitope \cite{Gupta2006},  we define the virus fitness $w$ as the sum of the contribution in each site of epitope B
\begin{equation}\label{eq:fitness}
w = A_0 + \sum_{\text{epitope B}} \delta w_i.
\end{equation}
We denote by $\Delta\Delta G_i^{\alpha\gamma}$ the free energy difference to substitute amino acid $\alpha$ to amino acid $\gamma$ at site $i$.  We investigated two versions of the virus fitness landscape.  The first is to define $\delta w_i$ as a linear function of the free energy difference of the substitution
\begin{equation}\label{eq:fitness1}
\delta w_i = A_1 \frac{\Delta\Delta G_i^{\alpha\gamma}}{\Delta\Delta G_\mathrm{crit}}.
\end{equation}
The second is to define $\delta w_i$ as a step function
\begin{equation}\label{eq:fitness2}
\delta w_i = A_2 H\left(\Delta\Delta G_i^{\alpha\gamma} - \Delta\Delta G_\mathrm{crit}\right)
\end{equation}
in which $H$ is the Heaviside step function. Illustrated in the simulation below, either definition of fitness is sufficient to explain the observed immune evasion of the H3N2 virus.

%

\subsection{Selection in the Epitope}
\label{sec:Selection}

Evolution of the H3N2 virus is driven jointly by neutral evolution and selection \cite{Koelle2006}. Neutral evolution may be ongoing in sites outside the epitopes. The high substitution rate in epitope B suggests that selection is the major factor shaping the pattern of evolution in that epitope \cite{Shih2007}. Shown in \ref{tab:history} and \ref{tab:vaccine} are the historical substitutions. The significantly increased ranks of free energy differences suggests the existence of selection by the immune pressure for substitutions that have increased the free energy difference $\Delta\Delta G$ and decreased the antibody binding constant. The immune selection is directional: certain types of amino acids such as charged ones were initially more likely to be added into the epitope B \cite{Pan2011a} because they maximally decreased the antibody binding constant as indicated in \ref{tab:rank}. The heterogeneity of the expected free energy difference of a random substitution in \ref{tab:site} shows that each site in epitope B has a specific weight with regard to immune escape.


\ref{tab:history} also illustrates that the immune selection did not necessarily pick the amino acid with the highest rank of $\Delta\Delta G$ as the substituting amino acid. Amino acids with moderate rank were introduced into epitope B even for the fixed substitution T155Y. Therefore the historical evolution did not simply substitute amino acids by maximizing the free energy differences in \ref{tab:ddG}. This phenomenon is possibly due to two causes. First, the virus fitness may be insensitive to the $\Delta\Delta G$ values, e.g. $A_1$ in \ref{eq:fitness1} may be small, or amino acid substitutions with large $\Delta\Delta G$ values may contribute equivalently to the fitness, as in \ref{eq:fitness2}. Second, only a small fraction of virus in one host is shed by the host and infects the next host, so the population size of propagated virus from one host is smaller by several orders of magnitude than the total virus population size in the same host. Additionally, a seasonal bottleneck exists in the influenza virus circulation \cite{Rambaut2008}. Both random mutation and small population sizes lead to dramatic randomness in the evolution. Consequently, the evolution of H3 hemagglutinin is not solely determined by maximizing the free energy differences in \ref{tab:ddG} and minimizing the antibody binding constant, even if the virus is under immune selection. Instead, randomness plays a key role in the H3N2 virus evolution.

\subsection{A Picture of the H3N2 Virus Evolution}
\label{sec:Evolution}


Selection depends on the fitness of each virus genotype that is quantified as a non-decreasing function of the free energy difference $\Delta\Delta G$. Moderate selection in epitope B requires that fitness improvement is limited when $\Delta\Delta G$ is large. One possibility is that the ratio $A_1/A_0$ in \ref{eq:fitness1} is small. Another is that the fitness takes the form of  \ref{eq:fitness2} in which all substitutions with $\Delta\Delta G > \Delta\Delta G_\mathrm{crit}$ have equal fitness.

The virus evolution is also affected by the genetic drift. Genetic drift is a term which captures the random component of evolution due to the large size of the phase space of possible substitutions relative to the single set of substitutions that lead to the highest viral fitness. The effect of genetic drift is quantitatively reflected in the fixation process of a new strain, as shown in the simulation below. A narrow bottleneck of virus propagation allows only a small fraction of the progeny to survive, imposing a notable probability that a favorable substitution is lost in the next generation. The effect of genetic drift is to increase the randomness in the virus evolution so that observed substitutions are based on chance in addition to the fitness of these substitutions.


To model the H3N2 evolution discussed above, we ran two Monte Carlo simulations of the influenza evolution model. A population of $N$ sequences of epitope B with 21 sites were created and initialized as the wildtype A/Aichi/2/1968 sequence. Here $N = 10^3$ to account for a narrow genetic bottleneck of hemagglutinin and for tractability of the simulation.  We iterated the simulation program for 5,000 generations or about five years to recreate a pattern of evolution similar to that in history and shown in \ref{tab:history}. The random substitution rate of H3 hemagglutinin is roughly $4.5 \times 10^{-6}$ amino acid substitution/site/generation \cite{Nobusawa2006}. We let the number of substitutions follow a Poisson distribution with mean $\lambda = 21 \times 4.5 \times 10^{-6} N = 9.5 \times 10^{-5} N$ and randomly assigned the substitution sites. The substituting amino acid at each substitution site was randomly picked from the remaining 19 amino acids proportional to the historical frequencies observed in hemagglutinin. The fitness $w$ in the first simulation was calculated for each sequence using \ref{eq:fitness1} with $A_0 = 100$ and $A_1 = 3$ and that in the second simulation was calculated for each sequence using \ref{eq:fitness2} with $A_0 = 100$, $A_2 = 9$, and $\Delta\Delta G_\mathrm{crit} = 1.42$ kcal/mol. Note that by choosing $A_1 = 3$ for the first simulation, a random substitution causes the expected fitness to change from 100 to 104.9, and by choosing $A_2 = 9$ for the second simulation, a random substitution changes the expected fitness from 100 to 105.0. The size of the progeny of each sequences equals the fitness $w$ of the sequence if $w > 0$, and equals 0 if $w \le 0$. The next generation of sequences was initialized by randomly sampling $N$ sequences from the progeny sequences.

\begin{figure}
\centering
\vspace{-0.5in}
\subfigure[]{
\includegraphics[width=4in]{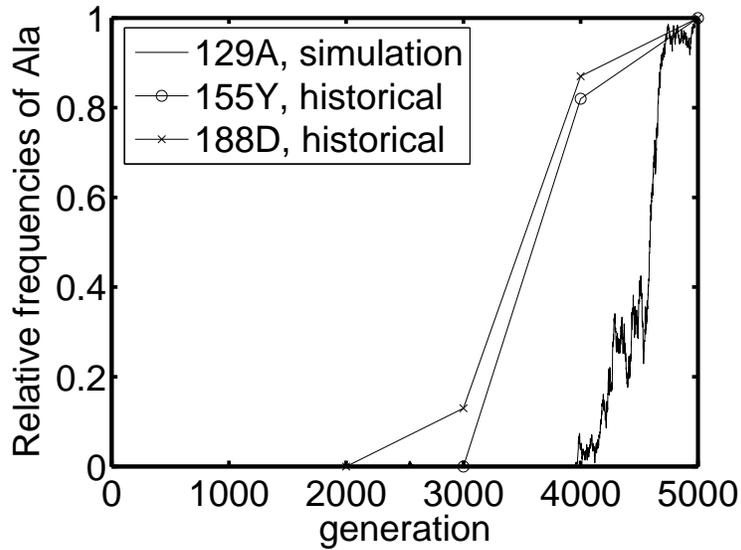}
\label{fig:G129A_method1}
}
\subfigure[]{
\includegraphics[width=4in]{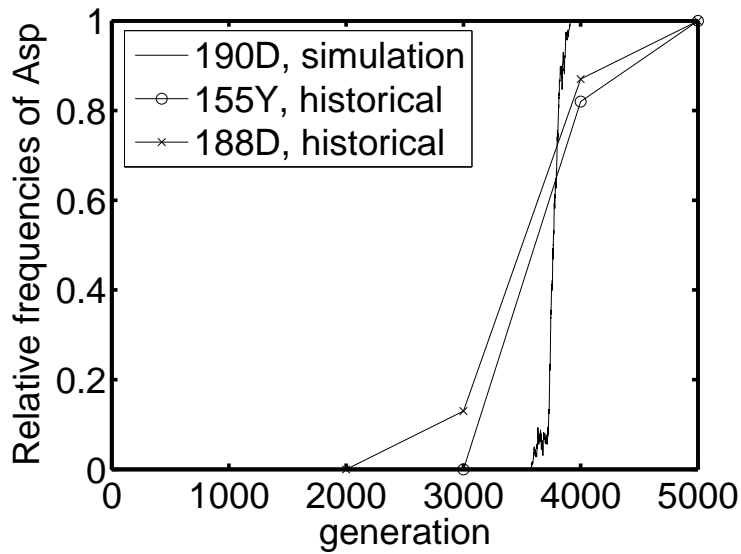}
\label{fig:E190D_method1}
}
\caption{Two fixed substitutions G129A and E190D generated by Monte Carlo simulation of epitope B using \ref{eq:fitness1}. Also plotted are two historical fixed substitutions in epitope B: T155Y fixed between 1971 and 1973, and N188D fixed between 1970 and 1973. The frequency data of historical substitutions are from Shih et al.\ \cite{Shih2007}. The origin of time axis is 1968. One thousand generation of the H3N2 virus is approximately one year. \ref{fig:G129A_method1} Substitution G129A causing the free energy difference $\Delta\Delta G = 3.33 \pm 0.29$ kcal/mol is fixed by the simulation. The rank of the free energy difference of G129A is 12 in 19 possible substitutions in site 129. \ref{fig:E190D_method1} Substitution E190D with $\Delta\Delta G = 18.75 \pm 0.32$ kcal/mol. The rank is 1 in 19 possible substitutions in site 190.}
\label{fig:simulation1}
\end{figure}

\begin{figure}
\centering
\vspace{-0.5in}
\subfigure[]{
\includegraphics[width=4in]{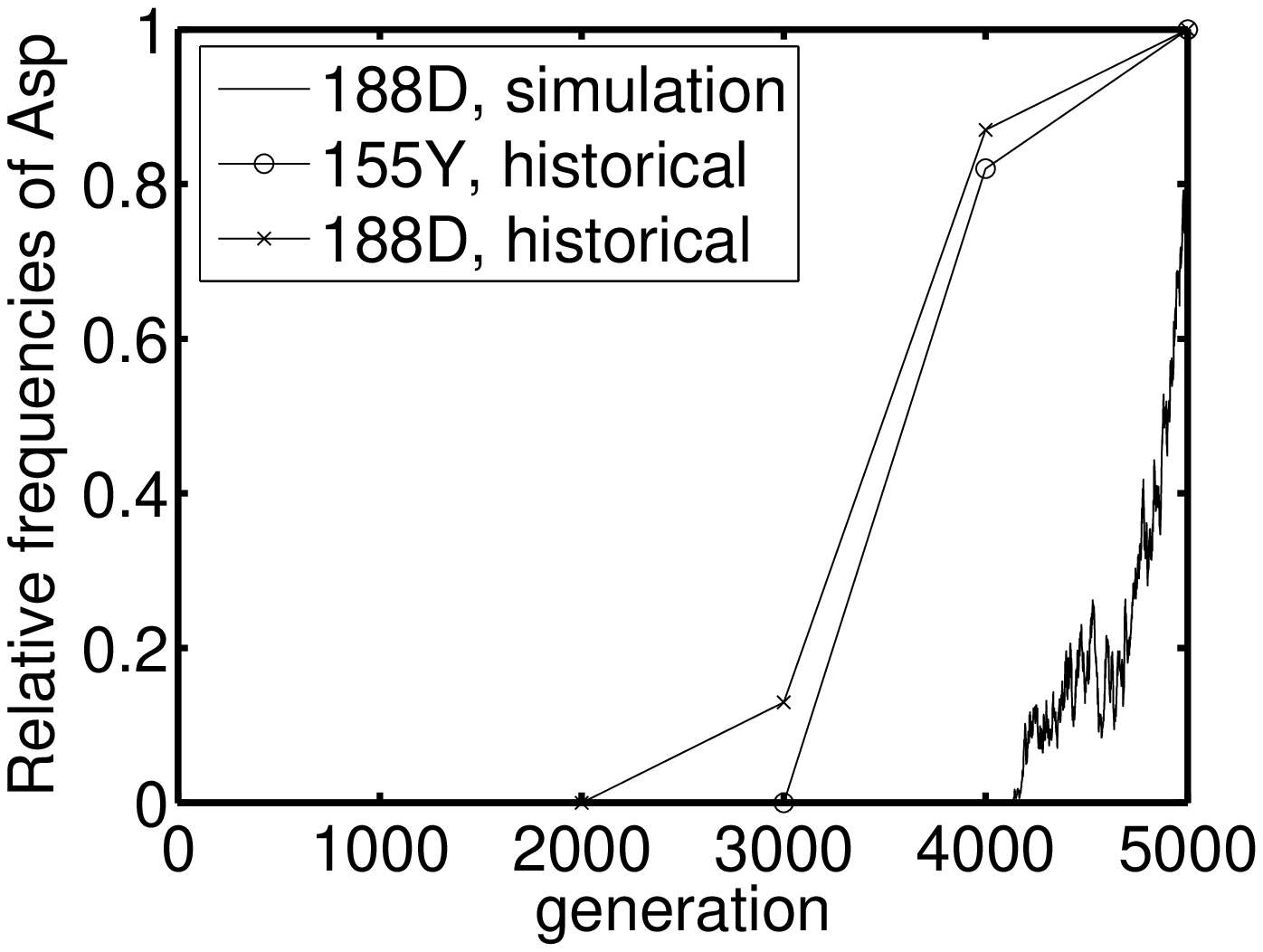}
\label{fig:N188D_method3}
}
\subfigure[]{
\includegraphics[width=4in]{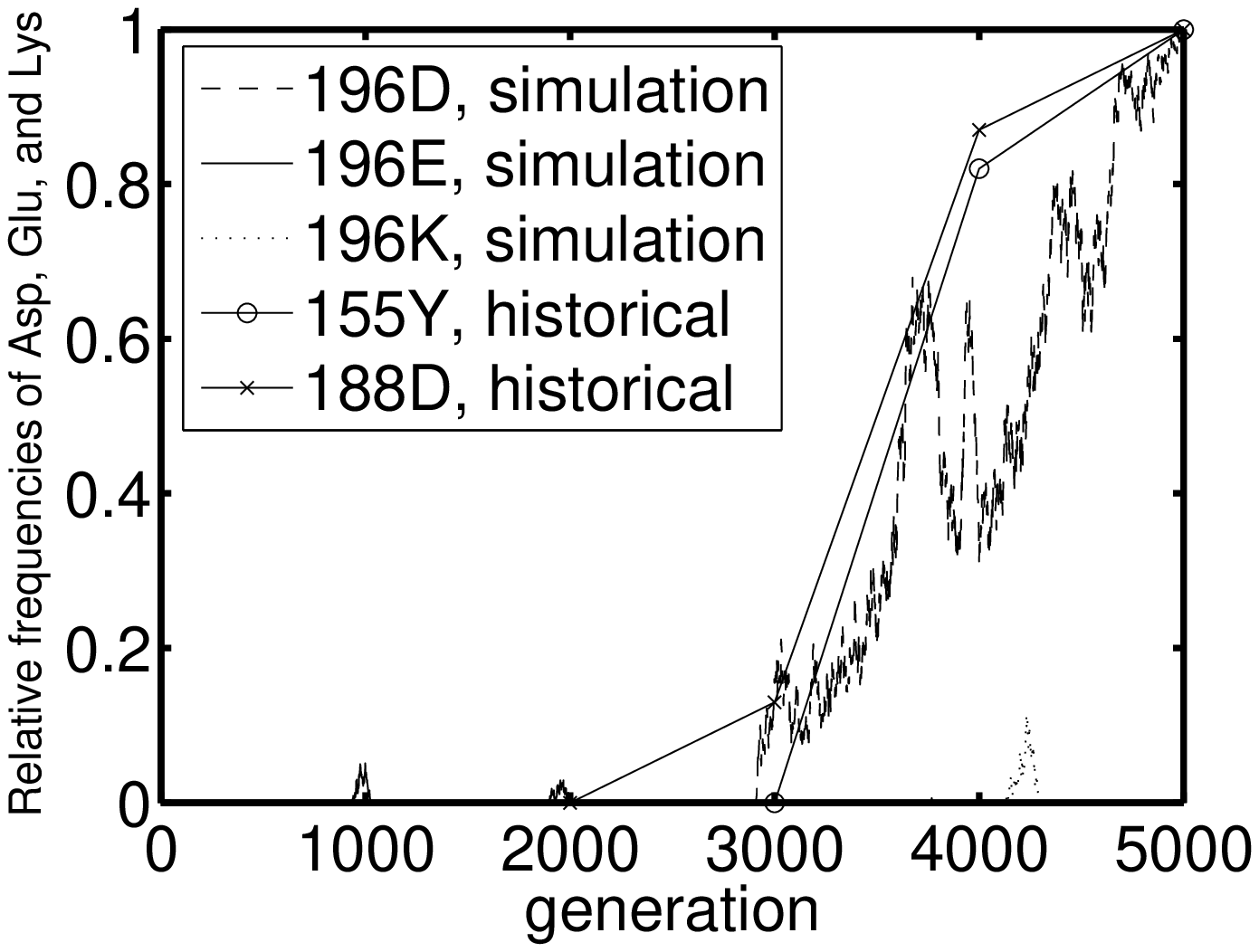}
\label{fig:V196D_method3}
}
\caption{Two fixed substitutions N188D and V196D generated by Monte Carlo simulation of epitope B using \ref{eq:fitness2}. Two historical fixed substitutions T155Y and N188D are also plotted, and data are from Shih et al.\ \cite{Shih2007}. \ref{fig:N188D_method3} Substitution N188D causing the free energy difference $\Delta\Delta G = 19.77 \pm 0.37$ kcal/mol is fixed by the simulation. The rank of the free energy difference of N188D is 1 in 19 possible substitutions in site 188. \ref{fig:V196D_method3} Substitution V196D with $\Delta\Delta G = 9.25 \pm 0.34$ kcal/mol. The rank is 5 in 19 possible substitutions in site 196. The proportions of substituting amino acids are represented by different line types.}
\label{fig:simulation2}
\end{figure}

The results of both simulations showed remarkable similarity to the observed substitutions in \ref{tab:history} with the bottleneck $N$ equal to $10^3$. See \ref{fig:simulation1} and \ref{fig:simulation2}. Amino acid substitutions generated in the simulation are usually distinct with those in \ref{tab:history} observed in history. The $\Delta\Delta G$ values of each substitution emerging in the simulation are nevertheless similar to those of the historical substitutions listed in \ref{tab:history}. As was observed in history in \ref{tab:history}, most of the substituted strains in the simulations with relative frequency greater than $1\%$ have positive $\Delta\Delta G$ values with the ranks of the substituting amino acids ranging from 1 to 10. The fixation of a newly emerged substitution takes about 1,000 generations or one year on average. Fixed substitutions mostly introduce amino acids with positive $\Delta\Delta G$ values in \ref{tab:ddG} and higher ranks in \ref{tab:rank}, and several of these fixed substitutions in simulation, such as E190D and N188D, have the highest $\Delta\Delta G$ values in the current site. However, fixed substitutions in the simulation are not always the substitutions with the highest $\Delta\Delta G$ values in \ref{tab:ddG}. These observations suggest that the Monte Carlo simulation considering the effect of substitution, selection, and genetic drift is able to reproduce the pattern of evolution observed in history. This simulation also shows that besides the free energy difference of each substitution, the mapping from the free energy landscape to the fitness landscape as well as the random genetic drift are dominant factors of the evolution in virus epitopes.

Shown in \ref{fig:simulation1} and \ref{fig:simulation2} for both simulations are the trajectories of relative frequencies of substituting amino acids. The trajectories are similar to historical observations of human H3N2 virus data \cite{Shih2007}.  For influenza, 1000 generations roughly equal one year.  The two substitutions T155Y and N188D were fixed in epitope B in 1968--1973.  As indicated by \ref{fig:simulation1} and \ref{fig:simulation2}, substitution T155Y emerged between generation 3000 and 4000, or equivalently between 1971 and 1972 from the emergence of the H3N2 virus in 1968 \cite{Shih2007}. Substitution T155Y was fixed between generation 4000 and 5000.  Similarly, substitution N188D emerged between generation 2000 and 3000 and was fixed between generation 4000 and 5000. The first simulation in which virus fitness is calculated using \ref{eq:fitness1} generated two fixed substitution, G129A that emerged at generation 4000 and was fixed by generation 5000, and E190D that emerged at generation 3600 and was fixed by generation 3900.  The second simulation using \ref{eq:fitness2} generated one fixed substitutions, V196D emerging at generation 2900 and fixed by generation 5000, and one substitution that nearly fixed, N188D emerging at generation 4100 and acquiring the relative frequency 0.84 at generation 5000.  The trajectories in both simulations resemble those of substitutions T155Y and N188D observed in history.  From these results, the two Monte Carlo simulations appear to capture the main factors of immune selection and genetic drift in evolution of the H3N2 virus.

\subsection{Multiple Substitutions}
\label{sec:Multiple}

In this work, we calculated the free energy difference for each possible substitution in epitope B. The free energy calculation for multiple substitutions is intractable using the current technology due to the combinatorially increasing calculation load for multiple substitutions. The issue of multiple substitutions is here addressed by assuming that the effect of immune evasion is well represented by the sum of the contribution in each substituted site in epitope B. Data indicate the independence of the immune evasion effect of the sites in epitope B \cite{Gupta2006}. We may, thus, assume that the free energy difference of the multiple substitution is the sum of the individual $\Delta\Delta G$ values available in \ref{tab:ddG} plus a minor correction term.


\subsection{Prediction of Future Virus Evolution}
\label{sec:Prediction}

The result of this work quantifies the reduction of the binding constant of antibody to virus for substitutions in epitope B with larger $\Delta\Delta G$ values and higher ranks of substituting amino acids. A newly emerging virus strain with larger antibody binding free energy difference has a greater probability to become the dominant strain in the next flu season. Note that due to random fluctuations in the large phase space of possible substitutions, actual trajectories deviate from the trajectory determined by choosing sites and substituting amino acids with greatest free energy differences. With a three dimensional structure of hemagglutinin of the current circulating virus and binding antibody, one is able to calculate the free energy landscape for all the possible single substitutions in the dominant epitope and estimate the \emph{a priori} escape probabilities in the next season. The dominant circulating influenza strain usually possesses amino acid substitutions from the vaccine strain against which memory antibodies are generated. Usually these substitutions disrupt the antibody binding process by decreasing the binding constant, as shown in \ref{tab:vaccine}. Thus one can predict vaccine effectiveness by evaluating the antibody binding constant against the dominant circulating strain, which is acquired by calculating free energy difference of the amino acid substitutions between the vaccine strain and the dominant circulating strain \cite{Gupta2006}. More accurate predictions of evolutionary pattern of virus as well as epidemiological data such as vaccine effectiveness may be obtained by optimally mapping the free energy landscape to the fitness landscape and taking into account random factors such as genetic drift in the evolution process.

\section{Conclusion}
\label{sec:Conclusion}

We introduced the Einstein crystal as a technology to improve the results of free energy calculation. By calculating the free energy difference of each amino acid substitution, we obtained the free energy landscape for substitutions in epitope B of hemagglutinin. There is notable variation between the values of free energy differences of different substitutions at different sites, because the identities of original and substituting amino acids, as well as the locations of amino acid substitutions, affect to differing degrees the antibody binding process. In this free energy landscape, we suggest that virus tends to evolve to higher $\Delta\Delta G$ values to escape binding of antibody. Counterbalancing this selection is random drift. Historical amino acid substitutions in epitope B and Monte Carlo simulations of the virus evolution using the free energy based virus fitness, in which random genetic drift of the virus adds statistical noise into the virus evolution process, showed that selected substitutions are biased to those with positive $\Delta\Delta G$ values.

\acknowledgement

This research was supported in part by the National Science Foundation through TeraGrid resources provided by Purdue and Indiana University under grant number TG-MCA05S015.  We specifically acknowledge the assistance of Phillip Cheeseman.  Part of the calculation is performed at computing resources at Research Computing Support Group, Rice University.  Keyao Pan's research was supported by a training fellowship from the Keck Center Nanobiology Training Program of the Gulf Coast Consortia (NIH Grant No.\ R90 DK071504).  This project was also partially supported by DARPA grant HR 00110910055.

\bibliography{achemso}

\providecommand*\mcitethebibliography{\thebibliography}
\csname @ifundefined\endcsname{endmcitethebibliography}
  {\let\endmcitethebibliography\endthebibliography}{}
\begin{mcitethebibliography}{51}
\providecommand*\natexlab[1]{#1}
\providecommand*\mciteSetBstSublistMode[1]{}
\providecommand*\mciteSetBstMaxWidthForm[2]{}
\providecommand*\mciteBstWouldAddEndPuncttrue
  {\def\EndOfBibitem{\unskip.}}
\providecommand*\mciteBstWouldAddEndPunctfalse
  {\let\EndOfBibitem\relax}
\providecommand*\mciteSetBstMidEndSepPunct[3]{}
\providecommand*\mciteSetBstSublistLabelBeginEnd[3]{}
\providecommand*\EndOfBibitem{}
\mciteSetBstSublistMode{f}
\mciteSetBstMaxWidthForm{subitem}{(\alph{mcitesubitemcount})}
\mciteSetBstSublistLabelBeginEnd
  {\mcitemaxwidthsubitemform\space}
  {\relax}
  {\relax}

\bibitem[WHO()]{WHO2009}
{W}orld {H}ealth {O}rganization {M}edia {C}entre Influenza Fact Sheet 211.
  \url{http://www.who.int/mediacentre/factsheets/fs211/en/index.html}, accessed
  on August 10, 2010\relax
\mciteBstWouldAddEndPuncttrue
\mciteSetBstMidEndSepPunct{\mcitedefaultmidpunct}
{\mcitedefaultendpunct}{\mcitedefaultseppunct}\relax
\EndOfBibitem
\bibitem[Janeway et~al.(2005)Janeway, Travers, Walport, and
  Shlomchik]{Janeway2005}
Janeway,~C.; Travers,~P.; Walport,~M.; Shlomchik,~M. \emph{Immunobiology: {T}he
  immune system in health and disease}, 6th ed.; Garland Science: New York,
  2005; p 430\relax
\mciteBstWouldAddEndPuncttrue
\mciteSetBstMidEndSepPunct{\mcitedefaultmidpunct}
{\mcitedefaultendpunct}{\mcitedefaultseppunct}\relax
\EndOfBibitem
\bibitem[Gupta et~al.(2006)Gupta, Earl, and Deem]{Gupta2006}
Gupta,~V.; Earl,~D.~J.; Deem,~M.~W. \emph{Vaccine} \textbf{2006}, \emph{24},
  3881--3888\relax
\mciteBstWouldAddEndPuncttrue
\mciteSetBstMidEndSepPunct{\mcitedefaultmidpunct}
{\mcitedefaultendpunct}{\mcitedefaultseppunct}\relax
\EndOfBibitem
\bibitem[Lee et~al.(2008)Lee, Ha, Simmons, de~Jong, Chau, Schumacher, Peng,
  McMichael, Farrar, Smith, Townsend, Askonas, Rowland-Jones, and
  Dong]{Lee2008}
Lee,~L. Y.~H.; Ha,~D. L.~A.; Simmons,~C.; de~Jong,~M.~D.; Chau,~N. V.~V.;
  Schumacher,~R.; Peng,~Y.~C.; McMichael,~A.~J.; Farrar,~J.~J.; Smith,~G.~L.;
  Townsend,~A. R.~M.; Askonas,~B.~A.; Rowland-Jones,~S.; Dong,~T. \emph{J Clin
  Invest} \textbf{2008}, \emph{118}, 3478--3490\relax
\mciteBstWouldAddEndPuncttrue
\mciteSetBstMidEndSepPunct{\mcitedefaultmidpunct}
{\mcitedefaultendpunct}{\mcitedefaultseppunct}\relax
\EndOfBibitem
\bibitem[Pan et~al.(2011)Pan, Subieta, and Deem]{Pan2011b}
Pan,~K.; Subieta,~K.~C.; Deem,~M.~W. \emph{Protein Eng., Des. Sel.}
  \textbf{2011}, \emph{24}, 291--299\relax
\mciteBstWouldAddEndPuncttrue
\mciteSetBstMidEndSepPunct{\mcitedefaultmidpunct}
{\mcitedefaultendpunct}{\mcitedefaultseppunct}\relax
\EndOfBibitem
\bibitem[Pan and Deem(2009)Pan, and Deem]{Pan2009}
Pan,~K.; Deem,~M.~W. \emph{Vaccine} \textbf{2009}, \emph{27}, 5033--5034\relax
\mciteBstWouldAddEndPuncttrue
\mciteSetBstMidEndSepPunct{\mcitedefaultmidpunct}
{\mcitedefaultendpunct}{\mcitedefaultseppunct}\relax
\EndOfBibitem
\bibitem[Ferguson et~al.(2003)Ferguson, Galvani, and Bush]{Ferguson2003}
Ferguson,~N.~M.; Galvani,~A.~P.; Bush,~R.~M. \emph{Nature} \textbf{2003},
  \emph{422}, 428--433\relax
\mciteBstWouldAddEndPuncttrue
\mciteSetBstMidEndSepPunct{\mcitedefaultmidpunct}
{\mcitedefaultendpunct}{\mcitedefaultseppunct}\relax
\EndOfBibitem
\bibitem[Deem and Pan(2009)Deem, and Pan]{Deem2009}
Deem,~M.~W.; Pan,~K. \emph{Protein Eng., Des. Sel.} \textbf{2009}, \emph{22},
  543--546\relax
\mciteBstWouldAddEndPuncttrue
\mciteSetBstMidEndSepPunct{\mcitedefaultmidpunct}
{\mcitedefaultendpunct}{\mcitedefaultseppunct}\relax
\EndOfBibitem
\bibitem[Zhou et~al.(2008)Zhou, Das, and Royyuru]{Zhou2008}
Zhou,~R.~H.; Das,~P.; Royyuru,~A.~K. \emph{J Phys Chem B} \textbf{2008},
  \emph{112}, 15813--15820\relax
\mciteBstWouldAddEndPuncttrue
\mciteSetBstMidEndSepPunct{\mcitedefaultmidpunct}
{\mcitedefaultendpunct}{\mcitedefaultseppunct}\relax
\EndOfBibitem
\bibitem[Brooks et~al.(1983)Brooks, Bruccoleri, Olafson, States, Swaminathan,
  and Karplus]{Brooks1983}
Brooks,~B.~R.; Bruccoleri,~R.~E.; Olafson,~B.~D.; States,~D.~J.;
  Swaminathan,~S.; Karplus,~M. \emph{J Comput Chem} \textbf{1983}, \emph{4},
  187--217\relax
\mciteBstWouldAddEndPuncttrue
\mciteSetBstMidEndSepPunct{\mcitedefaultmidpunct}
{\mcitedefaultendpunct}{\mcitedefaultseppunct}\relax
\EndOfBibitem
\bibitem[Frenkel and Smit(2002)Frenkel, and Smit]{Frenkel2002}
Frenkel,~D.; Smit,~B. \emph{Understanding molecular simulation: from algorithms
  to applications}, 2nd ed.; Academic Press: San Diego, 2002; p 168\relax
\mciteBstWouldAddEndPuncttrue
\mciteSetBstMidEndSepPunct{\mcitedefaultmidpunct}
{\mcitedefaultendpunct}{\mcitedefaultseppunct}\relax
\EndOfBibitem
\bibitem[Beveridge and DiCapua(1989)Beveridge, and DiCapua]{Beveridge1989}
Beveridge,~D.~L.; DiCapua,~F.~M. \emph{Annu Rev Biophys Biophys Chem}
  \textbf{1989}, \emph{18}, 431--492\relax
\mciteBstWouldAddEndPuncttrue
\mciteSetBstMidEndSepPunct{\mcitedefaultmidpunct}
{\mcitedefaultendpunct}{\mcitedefaultseppunct}\relax
\EndOfBibitem
\bibitem[Mezei and Beveridge(1986)Mezei, and Beveridge]{Mezei1986}
Mezei,~M.; Beveridge,~D.~L. \emph{Ann NY Acad Sci} \textbf{1986}, \emph{482},
  1--23\relax
\mciteBstWouldAddEndPuncttrue
\mciteSetBstMidEndSepPunct{\mcitedefaultmidpunct}
{\mcitedefaultendpunct}{\mcitedefaultseppunct}\relax
\EndOfBibitem
\bibitem[Cross(1986)]{Cross1986}
Cross,~A.~J. \emph{Ann NY Acad Sci} \textbf{1986}, \emph{482}, 89--90\relax
\mciteBstWouldAddEndPuncttrue
\mciteSetBstMidEndSepPunct{\mcitedefaultmidpunct}
{\mcitedefaultendpunct}{\mcitedefaultseppunct}\relax
\EndOfBibitem
\bibitem[Beutler et~al.(1994)Beutler, Mark, Vanschaik, Gerber, and
  Vangunsteren]{Beutler1994}
Beutler,~T.~C.; Mark,~A.~E.; Vanschaik,~R.~C.; Gerber,~P.~R.;
  Vangunsteren,~W.~F. \emph{Chem Phys Lett} \textbf{1994}, \emph{222},
  529--539\relax
\mciteBstWouldAddEndPuncttrue
\mciteSetBstMidEndSepPunct{\mcitedefaultmidpunct}
{\mcitedefaultendpunct}{\mcitedefaultseppunct}\relax
\EndOfBibitem
\bibitem[Zacharias et~al.(1994)Zacharias, Straatsma, and
  Mccammon]{Zacharias1994}
Zacharias,~M.; Straatsma,~T.~P.; Mccammon,~J.~A. \emph{J Chem Phys}
  \textbf{1994}, \emph{100}, 9025--9031\relax
\mciteBstWouldAddEndPuncttrue
\mciteSetBstMidEndSepPunct{\mcitedefaultmidpunct}
{\mcitedefaultendpunct}{\mcitedefaultseppunct}\relax
\EndOfBibitem
\bibitem[Boresch and Karplus(1999)Boresch, and Karplus]{Boresch1999}
Boresch,~S.; Karplus,~M. \emph{J Phys Chem A} \textbf{1999}, \emph{103},
  103--118\relax
\mciteBstWouldAddEndPuncttrue
\mciteSetBstMidEndSepPunct{\mcitedefaultmidpunct}
{\mcitedefaultendpunct}{\mcitedefaultseppunct}\relax
\EndOfBibitem
\bibitem[Boresch and Karplus(1999)Boresch, and Karplus]{Boresch1999b}
Boresch,~S.; Karplus,~M. \emph{J Phys Chem A} \textbf{1999}, \emph{103},
  119--136\relax
\mciteBstWouldAddEndPuncttrue
\mciteSetBstMidEndSepPunct{\mcitedefaultmidpunct}
{\mcitedefaultendpunct}{\mcitedefaultseppunct}\relax
\EndOfBibitem
\bibitem[Roux(1996)]{Roux1996}
Roux,~B. \emph{Biophys J} \textbf{1996}, \emph{71}, 3177--3185\relax
\mciteBstWouldAddEndPuncttrue
\mciteSetBstMidEndSepPunct{\mcitedefaultmidpunct}
{\mcitedefaultendpunct}{\mcitedefaultseppunct}\relax
\EndOfBibitem
\bibitem[Nina et~al.(1997)Nina, Beglov, and Roux]{Nina1997}
Nina,~M.; Beglov,~D.; Roux,~B. \emph{J Phys Chem B} \textbf{1997}, \emph{101},
  5239--5248\relax
\mciteBstWouldAddEndPuncttrue
\mciteSetBstMidEndSepPunct{\mcitedefaultmidpunct}
{\mcitedefaultendpunct}{\mcitedefaultseppunct}\relax
\EndOfBibitem
\bibitem[Essex et~al.(1997)Essex, Severance, TiradoRives, and
  Jorgensen]{Essex1997}
Essex,~J.~W.; Severance,~D.~L.; TiradoRives,~J.; Jorgensen,~W.~L. \emph{J Phys
  Chem B} \textbf{1997}, \emph{101}, 9663--9669\relax
\mciteBstWouldAddEndPuncttrue
\mciteSetBstMidEndSepPunct{\mcitedefaultmidpunct}
{\mcitedefaultendpunct}{\mcitedefaultseppunct}\relax
\EndOfBibitem
\bibitem[Price and Jorgensen(2001)Price, and Jorgensen]{Price2001}
Price,~D.~J.; Jorgensen,~W.~L. \emph{J Comput Aided Mol Des} \textbf{2001},
  \emph{15}, 681--695\relax
\mciteBstWouldAddEndPuncttrue
\mciteSetBstMidEndSepPunct{\mcitedefaultmidpunct}
{\mcitedefaultendpunct}{\mcitedefaultseppunct}\relax
\EndOfBibitem
\bibitem[Zacharias et~al.(1993)Zacharias, Straatsma, Mccammon, and
  Quiocho]{Zacharias1993}
Zacharias,~M.; Straatsma,~T.~P.; Mccammon,~J.~A.; Quiocho,~F.~A.
  \emph{Biochemistry} \textbf{1993}, \emph{32}, 7428--7434\relax
\mciteBstWouldAddEndPuncttrue
\mciteSetBstMidEndSepPunct{\mcitedefaultmidpunct}
{\mcitedefaultendpunct}{\mcitedefaultseppunct}\relax
\EndOfBibitem
\bibitem[Kaliman et~al.(2010)Kaliman, Nemukhin, and Varfolomeev]{Kaliman2010}
Kaliman,~I.; Nemukhin,~A.; Varfolomeev,~S. \emph{J Chem Theory Comput}
  \textbf{2010}, \emph{6}, 184--189\relax
\mciteBstWouldAddEndPuncttrue
\mciteSetBstMidEndSepPunct{\mcitedefaultmidpunct}
{\mcitedefaultendpunct}{\mcitedefaultseppunct}\relax
\EndOfBibitem
\bibitem[Crespo et~al.(2005)Crespo, Marti, Estrin, and Roitberg]{Crespo2005}
Crespo,~A.; Marti,~M.~A.; Estrin,~D.~A.; Roitberg,~A.~E. \emph{J Am Chem Soc}
  \textbf{2005}, \emph{127}, 6940--6941\relax
\mciteBstWouldAddEndPuncttrue
\mciteSetBstMidEndSepPunct{\mcitedefaultmidpunct}
{\mcitedefaultendpunct}{\mcitedefaultseppunct}\relax
\EndOfBibitem
\bibitem[Takahashi et~al.(2005)Takahashi, Kawashima, Nitta, and
  Matubayasi]{Takahashi2005}
Takahashi,~H.; Kawashima,~Y.; Nitta,~T.; Matubayasi,~N. \emph{J Chem Phys}
  \textbf{2005}, \emph{123}, 124504\relax
\mciteBstWouldAddEndPuncttrue
\mciteSetBstMidEndSepPunct{\mcitedefaultmidpunct}
{\mcitedefaultendpunct}{\mcitedefaultseppunct}\relax
\EndOfBibitem
\bibitem[Wang et~al.(2007)Wang, Hu, and Zhang]{Wang2007}
Wang,~S.~L.; Hu,~P.; Zhang,~Y.~K. \emph{J Phys Chem B} \textbf{2007},
  \emph{111}, 3758--3764\relax
\mciteBstWouldAddEndPuncttrue
\mciteSetBstMidEndSepPunct{\mcitedefaultmidpunct}
{\mcitedefaultendpunct}{\mcitedefaultseppunct}\relax
\EndOfBibitem
\bibitem[Deng and Roux(2006)Deng, and Roux]{Deng2006}
Deng,~Y.~Q.; Roux,~B. \emph{J Chem Theory Comput} \textbf{2006}, \emph{2},
  1255--1273\relax
\mciteBstWouldAddEndPuncttrue
\mciteSetBstMidEndSepPunct{\mcitedefaultmidpunct}
{\mcitedefaultendpunct}{\mcitedefaultseppunct}\relax
\EndOfBibitem
\bibitem[Frenkel and Ladd(1984)Frenkel, and Ladd]{Frenkel1984}
Frenkel,~D.; Ladd,~A. J.~C. \emph{J Chem Phys} \textbf{1984}, \emph{81},
  3188--3193\relax
\mciteBstWouldAddEndPuncttrue
\mciteSetBstMidEndSepPunct{\mcitedefaultmidpunct}
{\mcitedefaultendpunct}{\mcitedefaultseppunct}\relax
\EndOfBibitem
\bibitem[Noya et~al.(2008)Noya, Conde, and Vega]{Noya2008}
Noya,~E.~G.; Conde,~M.~M.; Vega,~C. \emph{J Chem Phys} \textbf{2008},
  \emph{129}, 104704\relax
\mciteBstWouldAddEndPuncttrue
\mciteSetBstMidEndSepPunct{\mcitedefaultmidpunct}
{\mcitedefaultendpunct}{\mcitedefaultseppunct}\relax
\EndOfBibitem
\bibitem[Frenkel and Smit(2002)Frenkel, and Smit]{Frenkel2002b}
Frenkel,~D.; Smit,~B. \emph{Understanding molecular simulation: from algorithms
  to applications}, 2nd ed.; Academic Press: San Diego, 2002; p 248\relax
\mciteBstWouldAddEndPuncttrue
\mciteSetBstMidEndSepPunct{\mcitedefaultmidpunct}
{\mcitedefaultendpunct}{\mcitedefaultseppunct}\relax
\EndOfBibitem
\bibitem[Meijer et~al.(1990)Meijer, Frenkel, Lesar, and Ladd]{Meijer1990}
Meijer,~E.~J.; Frenkel,~D.; Lesar,~R.~A.; Ladd,~A. J.~C. \emph{J Chem Phys}
  \textbf{1990}, \emph{92}, 7570--7575\relax
\mciteBstWouldAddEndPuncttrue
\mciteSetBstMidEndSepPunct{\mcitedefaultmidpunct}
{\mcitedefaultendpunct}{\mcitedefaultseppunct}\relax
\EndOfBibitem
\bibitem[Ryckaert et~al.(1977)Ryckaert, Ciccotti, and Berendsen]{Ryckaert1977}
Ryckaert,~J.~P.; Ciccotti,~G.; Berendsen,~H. J.~C. \emph{J Comput Phys}
  \textbf{1977}, \emph{23}, 327--341\relax
\mciteBstWouldAddEndPuncttrue
\mciteSetBstMidEndSepPunct{\mcitedefaultmidpunct}
{\mcitedefaultendpunct}{\mcitedefaultseppunct}\relax
\EndOfBibitem
\bibitem[Bennett(1975)]{Bennett1975}
Bennett,~C.~H. \emph{J Comput Phys} \textbf{1975}, \emph{19}, 267--279\relax
\mciteBstWouldAddEndPuncttrue
\mciteSetBstMidEndSepPunct{\mcitedefaultmidpunct}
{\mcitedefaultendpunct}{\mcitedefaultseppunct}\relax
\EndOfBibitem
\bibitem[Pomes and Mccammon(1990)Pomes, and Mccammon]{Pomes1990}
Pomes,~R.; Mccammon,~J.~A. \emph{Chem Phys Lett} \textbf{1990}, \emph{166},
  425--428\relax
\mciteBstWouldAddEndPuncttrue
\mciteSetBstMidEndSepPunct{\mcitedefaultmidpunct}
{\mcitedefaultendpunct}{\mcitedefaultseppunct}\relax
\EndOfBibitem
\bibitem[Feenstra et~al.(1999)Feenstra, Hess, and Berendsen]{Feenstra1999}
Feenstra,~K.~A.; Hess,~B.; Berendsen,~H. J.~C. \emph{J Comput Chem}
  \textbf{1999}, \emph{20}, 786--798\relax
\mciteBstWouldAddEndPuncttrue
\mciteSetBstMidEndSepPunct{\mcitedefaultmidpunct}
{\mcitedefaultendpunct}{\mcitedefaultseppunct}\relax
\EndOfBibitem
\bibitem[Rao et~al.(1987)Rao, Singh, Bash, and Kollman]{Rao1987}
Rao,~S.~N.; Singh,~U.~C.; Bash,~P.~A.; Kollman,~P.~A. \emph{Nature}
  \textbf{1987}, \emph{328}, 551--554\relax
\mciteBstWouldAddEndPuncttrue
\mciteSetBstMidEndSepPunct{\mcitedefaultmidpunct}
{\mcitedefaultendpunct}{\mcitedefaultseppunct}\relax
\EndOfBibitem
\bibitem[Flyvbjerg and Petersen(1989)Flyvbjerg, and Petersen]{Flyvbjerg1989}
Flyvbjerg,~H.; Petersen,~H.~G. \emph{J Chem Phys} \textbf{1989}, \emph{91},
  461--466\relax
\mciteBstWouldAddEndPuncttrue
\mciteSetBstMidEndSepPunct{\mcitedefaultmidpunct}
{\mcitedefaultendpunct}{\mcitedefaultseppunct}\relax
\EndOfBibitem
\bibitem[Morgan and Massi(2010)Morgan, and Massi]{Morgan2010}
Morgan,~B.~R.; Massi,~F. \emph{J Chem Theory Comput} \textbf{2010}, \emph{6},
  1884--1893\relax
\mciteBstWouldAddEndPuncttrue
\mciteSetBstMidEndSepPunct{\mcitedefaultmidpunct}
{\mcitedefaultendpunct}{\mcitedefaultseppunct}\relax
\EndOfBibitem
\bibitem[H\"unenberger and McCammon(1999)H\"unenberger, and
  McCammon]{Hunenberger1999}
H\"unenberger,~P.~H.; McCammon,~J.~A. \emph{J Chem Phys} \textbf{1999},
  \emph{110}, 1856--1872\relax
\mciteBstWouldAddEndPuncttrue
\mciteSetBstMidEndSepPunct{\mcitedefaultmidpunct}
{\mcitedefaultendpunct}{\mcitedefaultseppunct}\relax
\EndOfBibitem
\bibitem[Figueirido et~al.(1995)Figueirido, Delbuono, and Levy]{Figueirido1995}
Figueirido,~F.; Delbuono,~G.~S.; Levy,~R.~M. \emph{J Chem Phys} \textbf{1995},
  \emph{103}, 6133--6142\relax
\mciteBstWouldAddEndPuncttrue
\mciteSetBstMidEndSepPunct{\mcitedefaultmidpunct}
{\mcitedefaultendpunct}{\mcitedefaultseppunct}\relax
\EndOfBibitem
\bibitem[Pan et~al.(2011)Pan, Long, Sun, Tobin, Nara, and Deem]{Pan2011a}
Pan,~K.; Long,~J.; Sun,~H.; Tobin,~G.~J.; Nara,~P.~L.; Deem,~M.~W. \emph{J Mol
  Evol} \textbf{2011}, \emph{72}, 90--103\relax
\mciteBstWouldAddEndPuncttrue
\mciteSetBstMidEndSepPunct{\mcitedefaultmidpunct}
{\mcitedefaultendpunct}{\mcitedefaultseppunct}\relax
\EndOfBibitem
\bibitem[Sayle and Milnerwhite(1995)Sayle, and Milnerwhite]{Sayle1995}
Sayle,~R.~A.; Milnerwhite,~E.~J. \emph{Trends Biochem Sci} \textbf{1995},
  \emph{20}, 374--376\relax
\mciteBstWouldAddEndPuncttrue
\mciteSetBstMidEndSepPunct{\mcitedefaultmidpunct}
{\mcitedefaultendpunct}{\mcitedefaultseppunct}\relax
\EndOfBibitem
\bibitem[Pan and Deem()Pan, and Deem]{Pan_select}
Pan,~K.; Deem,~M.~W. An Entropy Method to Quantify Selection and Diversity in
  the Hemagglutinin of {H}3{N}2 Influenza. Submitted\relax
\mciteBstWouldAddEndPuncttrue
\mciteSetBstMidEndSepPunct{\mcitedefaultmidpunct}
{\mcitedefaultendpunct}{\mcitedefaultseppunct}\relax
\EndOfBibitem
\bibitem[NCB()]{NCBI_FLU}
NCBI Influenza Virus Resource.
  \url{http://www.ncbi.nlm.nih.gov/genomes/FLU/FLU.html}, accessed on August
  10, 2010\relax
\mciteBstWouldAddEndPuncttrue
\mciteSetBstMidEndSepPunct{\mcitedefaultmidpunct}
{\mcitedefaultendpunct}{\mcitedefaultseppunct}\relax
\EndOfBibitem
\bibitem[Smith et~al.(2004)Smith, Lapedes, de~Jong, Bestebroer, Rimmelzwaan,
  Osterhaus, and Fouchier]{Smith2004}
Smith,~D.~J.; Lapedes,~A.~S.; de~Jong,~J.~C.; Bestebroer,~T.~M.;
  Rimmelzwaan,~G.~F.; Osterhaus,~A. D. M.~E.; Fouchier,~R. A.~M. \emph{Science}
  \textbf{2004}, \emph{305}, 371--376\relax
\mciteBstWouldAddEndPuncttrue
\mciteSetBstMidEndSepPunct{\mcitedefaultmidpunct}
{\mcitedefaultendpunct}{\mcitedefaultseppunct}\relax
\EndOfBibitem
\bibitem[Shih et~al.(2007)Shih, Hsiao, Ho, and Li]{Shih2007}
Shih,~A.~C.; Hsiao,~T.~C.; Ho,~M.~S.; Li,~W.~H. \emph{Proc. Natl. Acad. Sci.
  USA} \textbf{2007}, \emph{104}, 6283--6288\relax
\mciteBstWouldAddEndPuncttrue
\mciteSetBstMidEndSepPunct{\mcitedefaultmidpunct}
{\mcitedefaultendpunct}{\mcitedefaultseppunct}\relax
\EndOfBibitem
\bibitem[Koelle et~al.(2006)Koelle, Cobey, Grenfell, and Pascual]{Koelle2006}
Koelle,~K.; Cobey,~S.; Grenfell,~B.; Pascual,~M. \emph{Science} \textbf{2006},
  \emph{314}, 1898--1903\relax
\mciteBstWouldAddEndPuncttrue
\mciteSetBstMidEndSepPunct{\mcitedefaultmidpunct}
{\mcitedefaultendpunct}{\mcitedefaultseppunct}\relax
\EndOfBibitem
\bibitem[Rambaut et~al.(2008)Rambaut, Pybus, Nelson, Viboud, Taubenberger, and
  Holmes]{Rambaut2008}
Rambaut,~A.; Pybus,~O.~G.; Nelson,~M.~I.; Viboud,~C.; Taubenberger,~J.~K.;
  Holmes,~E.~C. \emph{Nature} \textbf{2008}, \emph{453}, 615--U2\relax
\mciteBstWouldAddEndPuncttrue
\mciteSetBstMidEndSepPunct{\mcitedefaultmidpunct}
{\mcitedefaultendpunct}{\mcitedefaultseppunct}\relax
\EndOfBibitem
\bibitem[Nobusawa and Sato(2006)Nobusawa, and Sato]{Nobusawa2006}
Nobusawa,~E.; Sato,~K. \emph{J. Virol.} \textbf{2006}, \emph{80}, 3675--3678,
  In the amino acid level, the average mutation rate of influenza A virus is
  converted to $4.5 \times 10^{-6}$ amino acid
  substitution/site/generation.\relax
\mciteBstWouldAddEndPunctfalse
\mciteSetBstMidEndSepPunct{\mcitedefaultmidpunct}
{}{\mcitedefaultseppunct}\relax
\EndOfBibitem
\end{mcitethebibliography}

\end{document}